\documentclass[twocolumn]{article}

\usepackage[english]{babel}
\usepackage{csquotes}

\usepackage[letterpaper,top=2cm,bottom=2cm,left=2cm,right=2cm,marginparwidth=1.75cm]{geometry}

\usepackage{booktabs}
\usepackage{amsmath}
\usepackage{amssymb}
\usepackage{physics}
\usepackage{graphicx}
\usepackage{outlines}
\usepackage{hyperref}
\usepackage{cleveref}
\usepackage{xcolor}
\usepackage{enumitem}
\usepackage[normalem]{ulem} 
\usepackage{appendix}
\usepackage{siunitx}

\usepackage{float}
\usepackage{caption, setspace}
\usepackage{subcaption}
\usepackage{placeins}
\usepackage[backend=biber,style=nature]{biblatex}
\title{Enabling Biomolecular Simulations with Neural Network Potentials in GROMACS}
\author{
Lukas M\"ullender\footnote{Department of Applied Physics, Science for Life Laboratory \& Swedish e-Science Research Center, KTH Royal Institute of Technology, Stockholm, Sweden}
\and
Berk Hess\footnotemark[1]
\and
Erik Lindahl\footnotemark[1]
\footnote{Department of Biochemistry and Biophysics, Science for Life Laboratory, Stockholm University, Stockholm, Sweden}
\footnote{Department of Physics, Chemistry and Biology, Link\"oping University, Link\"oping, Sweden}
\footnote{Department Chemistry, University of Illinois at Urbana-Champaign, Urbana, IL, United States}
\textsuperscript{1}
}

\begin{document}
\maketitle
\footnotetext[1]{Email: erik.lindahl@dbb.se.su}

\textbf{ABSTRACT. Neural network potentials (NNPs) are rapidly changing the landscape of state-of-the-art molecular dynamics (MD) simulations. To make full use of this development, the community needs flexible, easy-to-use interfaces firmly integrated with existing methodologies. To address this, we here present an interface for hybrid machine learning/molecular mechanics (ML/MM) simulations implemented in the widely used MD code GROMACS. The interface enables NNPs trained in the PyTorch framework to contribute energies and forces during MD simulations, either for selected subsets or entire molecular systems. By defining a flexible set of model inputs and outputs, the interface is agnostic to specific NNP architectures and can accommodate a wide range of descriptor-based and message-passing models. In particular, the design integrates NNP inference seamlessly into the extensive GROMACS molecular simulation ecosystem, providing users with the capability to straightforwardly combine NNPs with existing advanced sampling and free energy workflows. We demonstrate the capabilities of the interface using several representative applications, including enhanced sampling of peptide torsional free energy landscapes, absolute solvation free energy calculations, and protein--ligand simulations. We also run performance benchmarks on water boxes for several different NNP architectures. Our interface is available in recent GROMACS releases, and we believe it will provide a practical foundation for incorporating machine learning potentials into production MD simulations of biomolecular systems.
}

\section{Introduction}
Since their development in the 1970s~\cite{warshelTheoreticalStudiesEnzymic1976a}, hybrid quantum mechanical/molecular mechanical (QM/MM) simulations have established themselves as a powerful computational tool, combining the enthalpic accuracy of electronic structure methods with the powerful sampling capabilities of molecular dynamics (MD) simulations~\cite{linQMMMWhat2007}. They provide an ideal trade-off between accuracy and efficiency especially in biomolecular systems, which are highly heterogeneous and often necessitate large solvent boxes best treated classically~\cite{sennQMMMMethods2009}.
However, QM/MM simulations remain computationally intensive as a result of the bottleneck of \textit{ab initio} QM calculations, even for QM regions of moderate size, making the treatment of many biological and chemical phenomena costly and time-consuming.

In recent years, the development of a class of machine learning (ML) models called machine learning interactomic potentials (MLIPs), or more specifically neural network potentials (NNPs), is promising to address this issue, by reproducing energies and forces with QM accuracy at a significantly reduced computational cost \cite{unkeMachineLearningForce2021,behlerFourGenerationsHighDimensional2021}. Since the seminal work of Behler and Parrinello~\cite{behlerGeneralizedNeuralNetworkRepresentation2007}, many different architectures have demonstrated excellent predictive accuracy on a wide range of available benchmarks, including ANI~\cite{smithANI1ExtensibleNeural2017,devereuxExtendingApplicabilityANI2020}, AIMNet~\cite{zubatyukAccurateTransferableMultitask2019,anstineAIMNet2NeuralNetwork2025}, MACE~\cite{batatiaMACEHigherOrder2023}, PhysNet~\cite{unkePhysNetNeuralNetwork2019}, NequIP~\cite{batznerE3equivariantGraphNeural2022} and others. 

Generally, applications of MD simulations with NNPs fall into two main classes, each with their own challenges: Full ML simulations, where the entire system is modeled by an NNP similar in spirit to \textit{ab initio} MD, and hybrid ML/MM simulations akin to QM/MM approaches.
The first has found use in biomolecular applications~\cite{unkeBiomolecularDynamicsMachinelearned2024,kovacsMACEOFFShortRangeTransferable2025,kabyldaMolecularSimulationsPretrained2025} and particularly in materials science, where they are now the de-facto standard due to the lack of broadly applicable empirical force fields~\cite{mishinMachinelearningInteratomicPotentials2021}. However, these simulations face challenges related to high compute and memory requirements, which requires scaling to multiple GPUs~\cite{musaelianLearningLocalEquivariant2023,fuchsChemtraindeployParallelScalable2025}, as well as the accurate incorporation of long-range interactions, which is an area of ongoing research~\cite{unkeSpookyNetLearningForce2021,yaoTensorMol01ModelChemistry2018,koFourthgenerationHighdimensionalNeural2021,shaiduIncorporatingLongrangeElectrostatics2024}. \\
In the context of biomolecular simulations, where widely applicable force fields and water models are available, employing a more expensive model for the description of e.g. solvent-solvent interactions which are ultimately of lesser interest might not be the most efficient use of computational resources. For this reason, it remains desirable to perform hybrid ML/MM simulations, where only a central region of interest is modeled with the NNP, while the rest of the system is treated with classical force fields. Such approaches have already shown promise for example in its application to conformational and binding free energies of protein--ligand complexes \cite{laheySimulatingProteinLigand2020,rufaChemicalAccuracyAlchemical2020}, alchemical free energy calculations \cite{rufaChemicalAccuracyAlchemical2020, sabaneszariquieyEnhancingProteinLigand2024, karwounopoulosInsightsChallengesCorrecting2024} and vibrational spectroscopy \cite{gasteggerMachineLearningSolvent2021,semelakAdvancingMultiscaleMolecular2025,zinovjevImprovedDescriptionEnvironment2025}. 
These works give rise to the nontrivial question of how to treat long-range atomic interactions across the ML--MM boundary. Many works have implemented a scheme analogous to mechanical embedding (ME)~\cite{laheySimulatingProteinLigand2020,rufaChemicalAccuracyAlchemical2020,eastmanOpenMM8Molecular2024,galvelisNNPMMAccelerating2023}, which is straightforward to implement with NNPs but known to perform insufficiently in many QM/MM applications~\cite{linQMMMWhat2007,sennQMMMMethods2009}. There, the state-of-the-art is represented by electrostatic embedding, which has been adapted to NNPs by the EMLE method~\cite{zinovjevElectrostaticEmbeddingMachine2023,zinovjevEmleengineFlexibleElectrostatic2024} and others~\cite{gasteggerMachineLearningSolvent2021,semelakAdvancingMultiscaleMolecular2025,pultarNeuralNetworkPotential2025}. 

When it comes to the practical application of this breadth of different competing approaches in high-performance MD simulation engines, there is a clear need for flexible software interfaces that can accommodate the architectural diversity, while keeping compatibility with existing workflows. While direct implementations and targeted optimizations exist for specific architectures~\cite{kovacsMACEOFFShortRangeTransferable2025,galvelisNNPMMAccelerating2023}, a more viable alternative is to rely on popular ML frameworks like PyTorch~\cite{paszkePyTorchImperativeStyle2019} or JAX~\cite{jax2018github}. This strategy has been taken by several implementations of NNP interfaces in popular MD codes like ASE \cite{hjorthlarsenAtomicSimulationEnvironment2017}, OpenMM \cite{eastmanOpenMM8Molecular2024}, LAMMPS~\cite{thompsonLAMMPSFlexibleSimulation2022} and Amber \cite{pickeringTorchANIAmberBridgingNeural2025}. Furthermore, packages that offer training capabilities for specific NNP architectures often also implement interfaces to MD engines of their choice, like MACE \cite{batatiaMACEHigherOrder2023}, DeePMD-kit \cite{zengDeePMDkitV3MultipleBackend2025} and chemtrain-deploy \cite{fuchsChemtraindeployParallelScalable2025}.

In this paper, we present a general and flexible interface for performing ML/MM simulations in the widely used molecular simulation toolkit GROMACS \cite{abrahamGROMACSHighPerformance2015}. Our interface, which we call \texttt{nnpot}, allows for NNPs trained within the popular PyTorch framework to contribute to the force calculation for an arbitrary subset of atoms in the system, or its entirety. It is actively maintained as part of the main GROMACS repository since its 2025 release, with the 2026 version adding significant updates to model compatibility and applicability. To be used with the interface, an arbitrary NNP only has to conform to a user-defined set of expected inputs and outputs, allowing for a high degree of flexibility. 
While this approach allows for straightforward customization, it might provide a somewhat high barrier of entry for users less familiar with machine learning frameworks, which is why we also provide a GitHub repository with an extensive how-to guide and several examples. We note that while our examples focus on popular NNP architectures like ANI and MACE, our interface is designed to be general and to work with any NNP that conforms to the API set by the interface. \\
After briefly reviewing the main concepts of NNPs and ML/MM simulations, as well as an overview of the implementation and main features of our interface, we showcase its capabilities on a range of different applications: To check the compatibility with enhanced sampling techniques available in GROMACS, we compute the torsional free energy profile of alanine dipeptide in water; next, we perform absolute alchemical free energy calculations to determine the solvation free energy of a selected subset of molecules from the FreeSolv database \cite{mobleyFreeSolvDatabaseExperimental2014}; to study the effects of the size of the ML region on protein-ligand binding, we consider the example case of catechol bound to the lysozyme L99A/M102Q mutant; lastly, we compare the performance scaling of a range of popular state-of-the-art NNPs. 

\begin{figure*}[ht!]
    \centering
    \includegraphics[width=\linewidth]{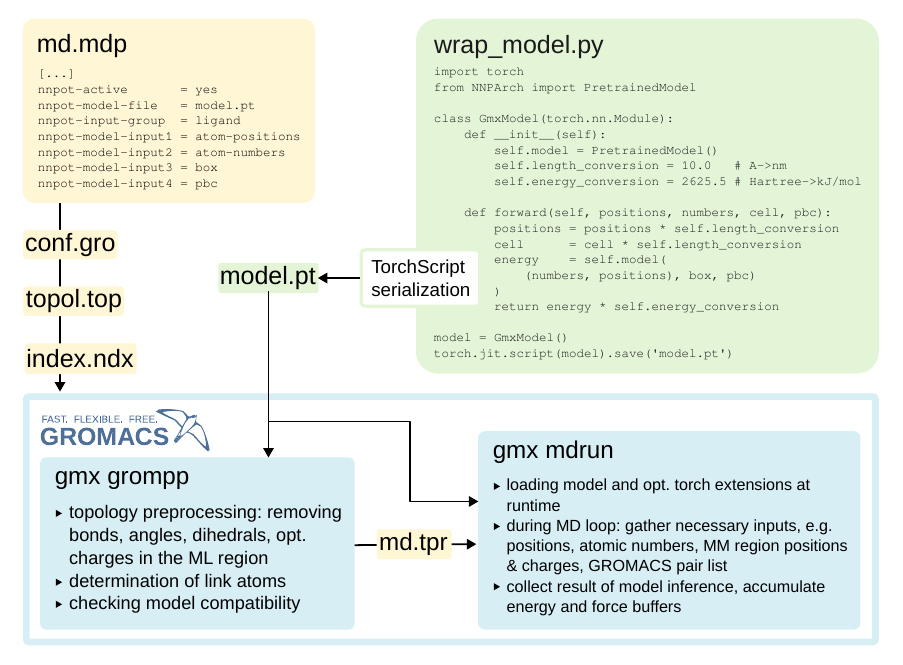}
    \caption{Schematic workflow of an MD simulation using the \texttt{nnpot} interface in GROMACS. The schematic shows the boilerplate code needed to wrap an example model \texttt{PretrainedModel} from an example \texttt{NNPArch} package for use with the interface. The model is exported via TorchScript and loaded within the interface at simulation runtime. Only minimal adjustments to existing simulation input files are needed.}
    \label{fig:nnpot}
\end{figure*}

\section{Theory}
\paragraph{Neural Network Potentials in ML/MM Simulations.}
In an ML/MM simulation, the total energy of the system can be expressed analogous to a QM/MM simulation as
\begin{equation}
    E_{tot} = E_{ML} + E_{MM} + E_{ML-MM},
\end{equation}
where $E_{ML}$ describes the energy of the ML subsystem predicted by the NNP, $E_{MM}$ the energy of the remainder of the system described at the MM level, and $E_{ML-MM}$ describes a coupling term to encompass all interactions between the ML and MM regions. \\
The energy $E_{ML}$, as predicted by the NNP, is a function of the atomic coordinates $\{\mathbf r_i^{ML}\}$ and nuclear charges $\{Z_i^{ML}\}$. To propagate the simulation, forces acting on the ML atoms also need to be inferred. While approaches have been proposed that predict forces directly \cite{chmielaMachineLearningAccurate2017}, a much more common approach is to derive them from the energy prediction to guarantee energy conservation by construction:
\begin{equation}
    \mathbf F_i^{ML} = - \nabla_i E_{ML},
    \label{eq:force}
\end{equation}
for a force acting on atom $i$ of the ML region. \\
For the description of the coupling term $E_{ML-MM}$, multiple possibilities exist, inspired by existing schemes from QM/MM approaches \cite{sennQMMMMethods2009}. The simplest of these is termed mechanical embedding, as it practically leaves all pairwise interactions between atoms in the ML and MM regions on the MM level:
\begin{equation}
    E_{ML-MM} = \sum_{i\in ML}\sum_{j\in MM} E_{ij}^{coul} + E_{ij}^{LJ},
\end{equation}
where $E_{ij}^{coul}$ and $E_{ij}^{LJ}$ are the Coulomb and Lennard-Jones interactions between particles $i$ and $j$, determined from fixed partial charges and Lennard-Jones parameters taken from the MM force field. \\
A more sophisticated strategy, and most widely used in QM/MM simulations, is the electrostatic embedding scheme, which can account for polarization effects on the ML region as a result of the surrounding MM atoms \cite{warshelTheoreticalStudiesEnzymic1976a}. While in QM/MM, this is done by including the MM atoms as point charges in the QM Hamiltonian, the implementation of such a scheme for NNPs is less straightforward.
A number of approaches proposed in the literature not only predict energies and forces on ML atoms, but also their \textit{in vacuo} partial charges, where the influence of the MM atoms is modeled by adding static and induced polarization corrections \cite{grassanoAssessmentEmbeddingSchemes2024,zinovjevElectrostaticEmbeddingMachine2023,semelakAdvancingMultiscaleMolecular2025,pultarNeuralNetworkPotential2025}. However, other schemes exist based on explicit electrostatic potential input to the NNP or the addition of a polarizable buffer region \cite{haghiriANIEFPModeling2024,lierBuRNNBufferRegion2022}. \\
Special attention needs to be given towards boundaries between ML and MM regions that cut through covalent bonds. In QM/MM simulations, a common strategy is to introduce a so-called link atom at an appropriate position along the QM--MM bond, which is only present in the QM subsystem and serves to complete the valence on the adjacent QM atom \cite{singhCombinedInitioQuantum1986,fieldDynamoLibraryMolecular2000}. A similar strategy can be employed in ML/MM simulations, where broken covalent bonds may lead to local chemical environments far outside the training regime of the NNP model. However, it should be noted that such a scheme is not without flaws and has to be applied with care, even in the case of QM/MM \cite{zlobinChallengesProteinQM2023}.

\section{Methods}
\paragraph{The \texttt{nnpot} Interface in GROMACS.}
The \texttt{nnpot} interface for performing ML/MM simulations in GROMACS is based on the modular MDModules framework, which greatly simplifies the calculation and integration of external forces on parts of the simulation system. The general workflow of such a simulation is illustrated in Fig. \ref{fig:nnpot}. It can be simply invoked by adding a new section \texttt{nnpot-active = yes} to the \texttt{.mdp} file during system preparation (see Fig.~\ref{fig:nnpot}). 
In its current form, it allows the user to specify a path to a NNP model that has been trained in PyTorch and exported via its TorchScript functionality. This enables us to take advantage of PyTorch's C++ API LibTorch and load models directly in GROMACS at runtime, without the added overhead of the Python interpreter. A similar approach is taken also by TorchANI-Amber \cite{pickeringTorchANIAmberBridgingNeural2025} and OpenMM/TorchForce~\cite{eastmanOpenMM8Molecular2024}. \\
Furthermore, multiple inputs to the model can be specified, such as atomic numbers, positions, the simulation box and type of periodic boundary conditions. The 2026 version adds the internal pair-list built by GROMACS, as well as the positions and charges of MM atoms as available inputs. 
The ML subsystem on which to apply the NNP is easily specified using an index group, which could be one of the default groups (such as \texttt{system} or \texttt{non-water}) or a custom index group supplied by a \texttt{.ndx} file. If such a selection leads to cut covalent bonds between ML and MM regions, virtual link atoms are introduced automatically, at a fixed distance from the ML atom along the bond to the ML atom, according to
\begin{equation}
    \mathbf r_L = \mathbf r_{ML} + d_{ML-L} \frac{\mathbf r_{MM} - \mathbf r_{ML}}{\left|\mathbf r_{MM} - \mathbf r_{ML}\right|}
\end{equation}
with link, ML and MM atom positions $\mathbf r_L$, $\mathbf r_{ML}$ and $\mathbf r_{MM}$, respectively. Link atom type and distance $d_{ML-L}$ can be specified by the user, defaulting to hydrogen and $d_{ML-L}=1$Å, respectively. The link atom is treated as a regular ML atom, and the resulting force on it redistributed according to the chain rule. \\
By itself, the interface currently only allows for simulations with the mechanical embedding (ME) scheme. However, with the addition of positions and charges on MM atoms as an available input in the 2026 version, it is possible to implement an electrostatic embedding (EE) scheme within the NNP model itself, as is done for example in the EMLE model~\cite{zinovjevEmleengineFlexibleElectrostatic2024}. In this case, the model can additionally add a force to the MM atoms, calculated via \textit{in vacuo} charges of the ML region additionally predicted by the NNP and used for evaluating the long-range electrostatic interactions, along with a polarization correction \cite{zinovjevElectrostaticEmbeddingMachine2023,semelakAdvancingMultiscaleMolecular2025}. If in use, the interface will automatically remove all classical partial charges of the ML region. \\
We also note that the model applied in the ML/MM simulation need not be a NNP in the usual sense, but could be used to apply e.g. a harmonic potential on every atom. In such cases, where the analytic form of the atomic gradients is known, forces can be supplied directly by the model to be used by the interface. 
For detailed information on the use and capabilities of the interface, we refer the user to the GROMACS reference manual \cite{ReferenceManualGROMACS}.

\paragraph{Simulation Details.}
\emph{Neural Network Potentials.}
All MD simulations in this paper were produced using GROMACS 2025.4 \cite{abrahamGROMACSHighPerformance2015} compiled with LibTorch (varying versions), with the exception of the simulations using MACE and EMLE models using a development version of GROMACS 2026. For the simulations using the ANI2x NNP, the corresponding pretrained model from the TorchANI package (v2.2.4) \cite{torchani} was used, in combination with the CUDA-optimized implementation of the NNPOps package (v0.6) \cite{galvelisNNPMMAccelerating2023}. For the performance benchmarks on water boxes, we additionally use the new version TorchANI 2.0 (v2.7.1)~\cite{pickeringTorchANI20Extensible2025}. For the simulations with MACE (v0.3.14), we use the small-size pre-trained foundation model MACE-OFF23 \cite{kovacsMACEOFFShortRangeTransferable2025}. Other NNP models used in this work are AIMNet2~\cite{anstineAIMNet2NeuralNetwork2025} (v0.1.0) and Nutmeg~\cite{eastmanNutmegSPICEModels2024}. The simulations with EMLE were performed using an ANI2x model wrapped with the electrostatic embedding model provided by the EMLE-engine~\cite{zinovjevEmleengineFlexibleElectrostatic2024}. For simplicity, we refer to this as the EMLE model throughout this paper. 

\emph{Enhanced Sampling.}
Enhanced sampling of the torsional free energy (FE) profile of alanine dipeptide in water was carried out using the Accelerated Weight Histogram method (AWH) \cite{lindahlAcceleratedWeightHistogram2014}. To prepare the system, an alanine dipeptide molecule was parametrized with the General Amber Force Field (GAFF) \cite{wangDevelopmentTestingGeneral2004} and solvated with 544 TIP3P \cite{jorgensenComparisonSimplePotential1983} water molecules. Energy minimization was performed for 10000 steps, after which the system was allowed to equilibrate for 100 ps in the NVT and 100 ps in the NPT ensemble using stochastic velocity and cell rescaling \cite{bussiCanonicalSamplingVelocity2007, bernettiPressureControlUsing2020} at 300 K and 1 bar, respectively, as well as a time step of 2 fs.
From the equilibrated state, enhanced sampling simulations with AWH were carried out at a time step of 1 fs for 10 ns with a force constant of 8000 kJ mol$^{-1}$ rad$^{-2}$, a diffusion constant of 5 rad$^2$/ps and estimated initial error of 10 kJ/mol, biasing the two dihedral angles $\phi$ and $\psi$ of the peptide. FE profiles were reconstructed using the \texttt{gmx awh} command line tool and filtered with a gaussian kernel of width 1.5$^\circ$.

\emph{Absolute Free Energy Calculations.}
Free energy calculations for the solvation of small molecules from the FreeSolv database \cite{mobleyFreeSolvDatabaseExperimental2014} were carried out using the free energy $\lambda$ mechanism in GROMACS. 
The systems were prepared using the topologies provided in the database and solvated with TIP3P water molecules. The systems were then minimized using a steepest descents algorithm for 5000 steps or until converged. Equilibration in the $NPT$ ensemble was performed in a single simulation for 100 ps, using a Langevin thermostat with a temperature of 298.15 K and \verb|c-rescale| pressure coupling at 1.013 bar.
All simulations were performed using a time step of 1 fs due to the use of NNPs. 
Production runs of 2 ns for each $\lambda$ were carried out in the $NPT$ ensemble starting with the fully coupled state, and performing calculations for successive values of $\lambda$ starting from the last frame of the previous simulation. In this way, simulations for the different lambda values cannot be parallelized, but no separate minimization and equilibration runs at each $\lambda$ are necessary. Instead, the first 10 ps of each production simulation were discarded from further analysis. 
Free energy calculations were carried out using the BAR \cite{bennettEfficientEstimationFree1976} implementation of the \verb|gmx bar| command line tool in GROMACS. 
To avoid issues with overlapping charges as interactions are successively turned off, we decouple Coulomb interactions before LJ interactions, and use a Beutler soft-core LJ potential.
The parameters of the thermostat and barostat, as well as the spacing of the $\lambda$ windows, are chosen to match those used in the reference calculations in the FreeSolv database, to ensure comparability.
The procedure as described above was applied uniformly to 30 molecules selected from the FreeSolv database without any per-case adjustments. Among them 20 molecules were selected at random, 5 were selected that had the largest discrepancy between experimental and calculated value reported in the data base, and 5 with the smallest discrepancy. This was done to check whether the use of NNPs can improve the calculated SFE values especially in cases where the classical FF fails, without sacrificing accuracy in cases that are well reproduced with classical FFs. 4 molecules, where the described procedure led to errors during system preparation, were discarded and replaced with new selections preserving the same criteria. The full selected dataset with all experimental and calculated values, as well as selection criteria, is shown in Table \ref{tab:sfe_results}.

\emph{Protein-Ligand binding Simulations}
The system of lysozyme L99A/M102Q bound to catechol (PDB entry 1XEP) was prepared using the CHARMM36/CGenFF force field \cite{vanommeslaegheCHARMMGeneralForce2010, bestOptimizationAdditiveCHARMM2012}. 10084 TIP3P water molecules and 8 chloride ions were added for solvation and neutralization. System minimization and equilibration was again carried out in an pure MM description as described above, with the addition of position restraints for the protein and ligand atoms during equilibration.
From the equilibrated configuration, 4 different simulations were started, one classical MM simulation and three ML/MM simulations, with the ML region comprising: 1. only the ligand in ME; 2. the ligand and the surrounding sidechains of protein residues (within 3 Å of the ligand) in ME; 3. only the ligand in EE. For the latter, link atoms were introduced, capping cut covalent bonds between $C_\alpha$ and $C_\beta$ atoms with a hydrogen atom at a distance of 1 Å along the bond. Again, we use the NNPOps-optimized version of ANI2x, and EMLE as the electrostatic embedding model. Protein-ligand interaction fingerprints were generated with the ProLIF package \cite{bouyssetProLIFLibraryEncode2021}.

\section{Results and Discussion}

\paragraph{Efficient Conformational Enhanced Sampling.}
\begin{figure}[ht!]
    \centering
    \includegraphics[width=\linewidth]{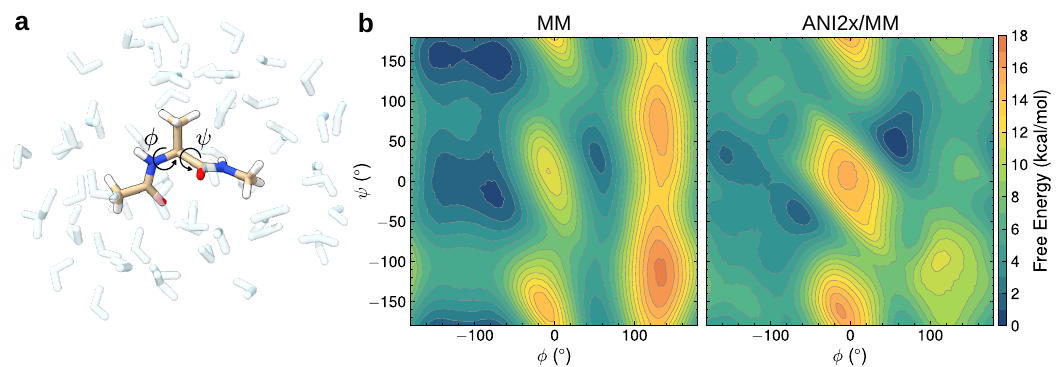}
    \caption{Conformational enhanced sampling with AWH. (a) Structure of an alanine dipeptide molecule solvated in water. (b) Ramachandran plot of the torsional free energy surface of alanine dipeptide in water, calculated using the AWH method for enhanced sampling biasing the two dihedral angles $\phi$ and $\psi$. The left side shows the reference calculated at MM level, while the right side shows results with the alanine dipeptide treated with ANI2x.}
    \label{fig:alanine}
\end{figure}

The conformational landscape of alanine dipeptide (Fig.~\ref{fig:alanine}a) has been studied long and widely in the literature as a model system for the torsional changes in the backbone of proteins~\cite{madisonSolventdependentConformationalDistributions1980,tobiasConformationalEquilibriumAlanine1992}, and used in the parametrization and validation of classical force fields~\cite{tianFf19SBAminoAcidSpecificProtein2020} and NNP architectures and implementations~\cite{zinovjevEmleengineFlexibleElectrostatic2024, semelakAdvancingMultiscaleMolecular2025, pultarNeuralNetworkPotential2025} alike.
Here, we report the Ramachandran plot of the torsional free energy profile of alanine dipeptide along its dihedral angles $\phi$ and $\psi$ in Fig.~\ref{fig:alanine}a. Our results reproduce important secondary structure minima consistently between ML/MM and pure MM simulations, and are in good agreement with those reported previously, both at the ML/MM level and a QM/MM reference at $\omega$B98x/6-31G* level of theory~\cite{zinovjevElectrostaticEmbeddingMachine2023,semelakAdvancingMultiscaleMolecular2025}. \\
As a key difference, our results were produced using the AWH method, which has been shown to be highly flexible and effective in calculating free energies in a number of different settings \cite{lindahlAcceleratedWeightHistogram2014,lundborgSkinPermeabilityPrediction2022}. This demonstrates that our NNP implementation can be readily combined with other sampling methodologies available in GROMACS. Furthermore, because of its efficient implementation, calculating free energies with AWH comes at very little additional cost compared to unbiased ML/MM simulations. In concrete terms, while an MM AWH simulation ran at 941 ns/day, a simulation using ANI2x for the description of the alanine dipeptide ran 42 ns/day, all at a timestep of 1 fs. In comparison, an unbiased ML/MM simulation ran at 51 ns/day.

\paragraph{Solubility of small molecules at the ML/MM level}
\begin{figure}[ht!]
    \centering
    \includegraphics[width=\linewidth]{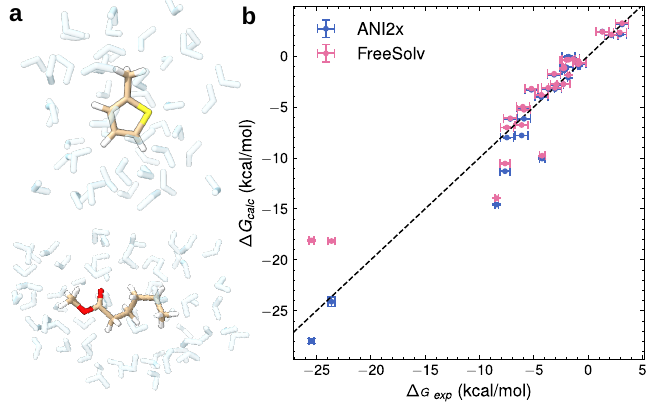}
    \caption{Solubility of small molecules at the ML/MM level. (a) Example molecules from the database, with surrounding water molecules: 2-methylthiophene (top) and methylhexanoate (bottom). (b) Solvation free energies for a subset of small molecules from the FreeSolv database, with inset showing small values of $|\Delta G_{solv}|$.}
    \label{fig:freesolv}
\end{figure}

In addition to conformational enhanced sampling methods, our ML/MM implementation also readily lends itself to alchemical free energy calculations. We showcase this capability by performing SFE calculations on a subset of 30 selected molecules from the FreeSolv database. The computation of SFEs is an important benchmark to gauge our ability to accurately model the binding interactions of proteins and small molecular ligands. Furthermore, they offer a quantitative comparison to experiment, which is critical to assess the quality of our molecular models. 
In Ref. \cite{karwounopoulosInsightsChallengesCorrecting2024}, the authors perform a systematic study of SFEs on a large subset of the FreeSolv database \cite{mobleyFreeSolvDatabaseExperimental2014}, using non-equilibrium switching and the ANI2x NNP to compute end-state corrections to SFE results derived from MM simulations. Here, we instead use an approach more similar to that of Ref. \cite{mobleyFreeSolvDatabaseExperimental2014}, simulating the entire decoupling process in an ML/MM setup and concentrate on a smaller, 30-molecule subset of the original database. 

The SFE results on our 30-molecule subset are illustrated in Fig. \ref{fig:freesolv}, where we compare the values calculated with the ANI2x NNP for the molecule with the experimental results found in the database and the reference calculations carried out using the GAFF force field parameters using AM1-BCC partial charges. All numerical values for all molecules are reported in the SI, Tab.~\ref{tab:sfe_results}.
In general, the mean absolute error for the ANI results score a mean absolute error (MAE) w.r.t. the experimental results of 1.20 kcal/mol, in contrast to 1.47 kcal/mol for the FreeSolv calculations. In terms of Pearson correlation coefficient, the ANI results score a value of 0.97, the FreeSolv results for the same molecules 0.92. It should be stated however, that these values present a biased measure of the performance of the two methods, because of the above mentioned non-stochastic selection methods. 

Nevertheless, these results suggest that the use of ANI2x can lead to a better estimation of the solvation free energy in some cases compared to GAFF. Crucially, it does not lead to a marked increase in the estimation error for any of the tested molecules. It is important to note that any observed improvement is due solely to an improved description of the intramolecular interactions as the solute-solvent interactions are still treated classically. 
We also note that in the cases where the reference calculations had the largest errors, the use of the NNP either led to the biggest observed improvements, or none at all. This points to different sources of the prediction error in these cases: where the parametrization of the intramolecular interactions was suboptimal, the NNP is able to significantly increase the prediction quality. In cases where it doesn't, it can be postulated that insufficiently modeled solute-solvent interactions, i.e. partial charges, are the cause of the error. 
While the authors in Ref. \cite{karwounopoulosInsightsChallengesCorrecting2024} report no statistically significant improvements of the overall prediction quality when employing end-state corrections with ANI2x, we cannot definitively corroborate or contradict their findings, as a systematic study is outside of the scope of this work. 

We note that in its current form, our NNP interface is only able to perform absolute alchemical free energy calculations. The calculation of relative free energies, where some atoms or entire chemical moieties are transformed into another, would require a more sophisticated single or dual-topology scheme, which is non-trivial to implement in the presence of an NNP.
Further improvements in the accuracy of the calculated SFE values could be attained by using a more accurate electrostatic embedding scheme like EMLE, which would require integrating the $\lambda$-parameter into the total electrostatic energy calculation of the model. 
Another possibility would be to describe the system entirely within the NNP, where special care has to be taken that the NNP accurately models long-range electrostatic interactions within itself \cite{koFourthgenerationHighdimensionalNeural2021}, and is able to appropriately describe the annihilation or creation of atoms \cite{harrymooreComputingSolvationFree2026}.

\paragraph{Effects of ML Region Size and Embedding on Protein-Ligand Binding}
\begin{figure*}[ht!]
    \centering
    \includegraphics[width=\linewidth]{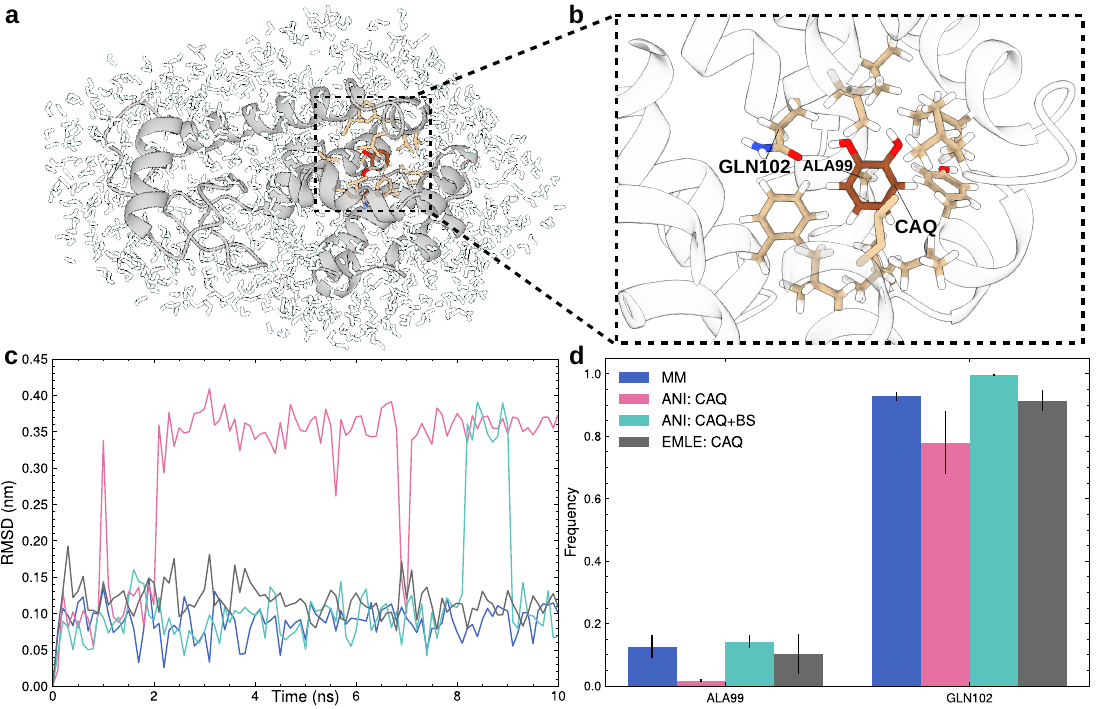}
    \caption{ML region size and embedding scheme affect protein-ligand binding in lysozyme. (a) Structure of lysozyme L99A/M102Q mutant bound to catechol, PDB ID 1XEP, with surrounding water molecules. (b) Close up of the binding site of catechol (CAQ, in sienna), showing the ligand and residues within 3 Å of it (in tan). (c) RMSD calculated with respect to the equilibrated starting structure for one pure MM and three ML/MM simulations: Using ANI2x for the catechol ligand (CAQ), ligand plus binding site (CAQ+BS), and EMLE for catechol. (d) Frequency of hydrogen bond formation between the ligand and two residues of the binding site, ALA99 and GLN102. The figure reports mean values and standard deviations over 3 repeats.}
    \label{fig:lysozyme}
\end{figure*}

In the following, we showcase the applicability of our interface to ML/MM simulations of protein-ligand complexes \cite{laheySimulatingProteinLigand2020}. As a test case, we pick the lysozyme L99A/M102Q mutant bound to catechol (see Fig. \ref{fig:lysozyme}a) because of its polar and well-characterized binding site \cite{boycePredictingLigandBinding2009}. We first perform a simulation at the MM level, as well as an ML/MM simulation with the ML region including only the ligand (CAQ). To test the automatic link atom scheme, we then also consider an ML region additionally containing amino acid sidechains within 3 Å of the ligand, with the ML/MM boundary bisecting the $C_\alpha-C_\beta$ bonds of the included residues (CAQ+BS). This setup can also be considered a proxy for the study of enzymatic reactions, where it is often necessary to include residues of the catalytic site in the QM(ML) region~\cite{kulikHowLargeShould2016}. 

In Fig. \ref{fig:lysozyme}c, we show the root mean square deviation (RMSD) of the ligand heavy atoms with respect to the initial configuration, calculated over 10 ns of simulation at the aforementioned levels of description. Results for all three performed replicates are shown in Fig.~\ref{fig:rmsd_all}. It is apparent that at the MM level, the initial ligand binding pose is stable over the course of the simulation. However, switching to ANI2x for the description of the ligand causes it to preferentially assume a different binding mode. Interestingly, the addition of the surrounding binding side residues into the ML region seems to re-stabilize the initial pose. This is further confirmed by comparing the frequency of hydrogen bonding interactions between the ligand and protein residues of the binding site, ALA99 and GLN102. We observe a drop in hydrogen bonding frequency for the CAQ simulations, which are restored in the CAQ+BS case. To assess statistical significance of the results, we analyzed results over 3 repeats of the simulations. We note that for the ALA99 residue, the hydrogen bond acceptor is the carbonyl oxygen of the backbone, which is not part of the ML region.
As we expect the inclusion of more interactions in the NNP for the CAQ+BS case to be more accurate, this result could be indicative of the limitations of the mechanical embedding scheme used in this application. Empirical FFs are parametrized with bonded and non-bonded parameters to reproduce target observables in conjunction, and simply replacing either of them with a different method is in principle invalid and can be detrimental.

To explore this more closely, we also performed simulations with only the CAQ input to the NNP, but additionally employing the EMLE model~\cite{zinovjevEmleengineFlexibleElectrostatic2024} for the ML-MM interactions, which has shown promise in a number of applications~\cite{zinovjevElectrostaticEmbeddingMachine2023, zinovjevEmleengineFlexibleElectrostatic2024, zinovjevImprovedDescriptionEnvironment2025}. Again, we observe hydrogen bonding frequencies of the same magnitude as in the MM case. This could be the effect of stronger interactions between the ligand and binding pocket when accounting for polarization effects on the ML region. A similar effect can be seen when studying solute-solvent effects in an ML/MM simulation of a phenol molecule in water, where previous studies have shown that an electrostatic embedding is necessary to accurately reproduce QM/MM results~\cite{semelakAdvancingMultiscaleMolecular2025}. We here additionally confirm that the EMLE model is able to reproduce this result, with the corresponding analysis shown in Fig.~\ref{fig:phenol}. Taken together, these results suggest that the destabilization of the binding pose could be an artifact of the mechanical embedding scheme. 

\begin{figure*}[ht!]
    \centering
    \includegraphics[width=\linewidth]{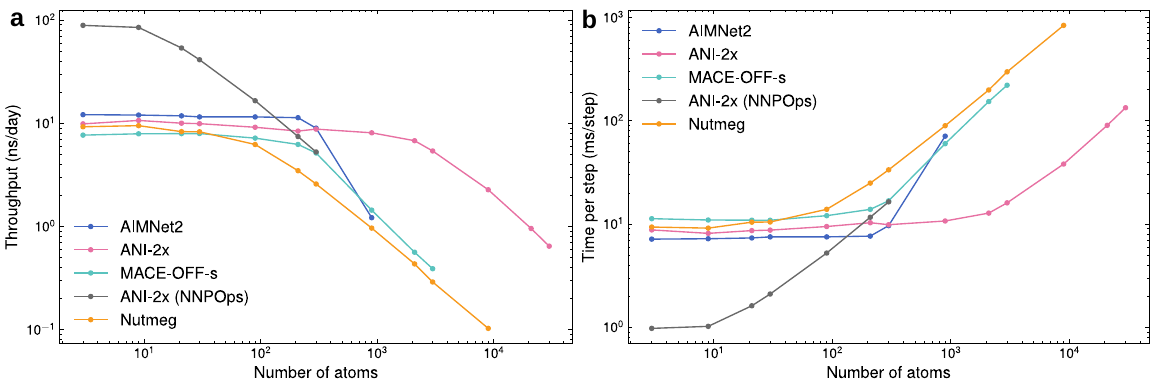}
    \caption{Performance scaling of different NNP Architectures on bulk water. Results shown are averages over one 3000 step MD simulation, resetting performance counters after 500 steps. Where no performance data is shown, this indicates that the calculations exceeded GPU memory for these system sizes. The figure shows (a) the throughput in terms of simulated time per wall time in units of ns/day at a 1 fs timestep, and (b) the performance in terms of walltime per simulation step in ms. Results were obtained on a NVIDIA RTX 3070 GPU at 32-bit floating point precision. }
    \label{fig:performance}
\end{figure*}

Thus, when simulating protein-ligand systems, special care should be taken when selecting the molecules and residues to include in the ML region. Due to the excellent scaling of NNP models, increasing the size of the ML region comes at a comparatively modest increase in computational cost. In this case, ML/MM simulations including only the ligand (14 ML atoms) ran at a performance of 58 ns/day, adding the side-chains of residues of the binding site (107 ML atoms, including link atoms) ran at 13 ns/day, and the EMLE model (14 ML atoms) at 9 ns/day, all at a time step of 1 fs. For reference, the MM simulation in GROMACS ran at a performance of 558 ns/day at a time step of 2 fs.

\paragraph{Performance Scaling of Different NNP Architectures}
Lastly, we focus on the computational performance of different NNP architectures when scaling to bulk water systems of different sizes. In contrast to other machine learning applications, a particular requirement for NNPs is low-latency single point inference, to be efficiently used in high-throughput MD simulations and consequently enable longer sampling times. Therefore, it is important for NNP architectures to be designed with both computational and memory efficiency in mind. While scaling NNP inference to multiple GPUs is feasible~\cite{musaelianLearningLocalEquivariant2023,fuchsChemtraindeployParallelScalable2025}, we consider it less relevant for hybrid ML/MM simulations, where ML regions are less likely to exceed GPU memory limits, especially with computational hardware rapidly advancing to accommodate the memory requirements of modern large-language models. Thus, we here only focus on single-device inference. 

Fig.~\ref{fig:performance} shows the results of our performance scaling experiments. We compare a number of popular state-of-the-art NNPs with available pretrained models: AIMNet2~\cite{anstineAIMNet2NeuralNetwork2025}, ANI2x (version 2.0)~\cite{devereuxExtendingApplicabilityANI2020,pickeringTorchANI20Extensible2025}, MACE-OFF-small~\cite{batatiaMACEHigherOrder2023,kovacsMACEOFFShortRangeTransferable2025} and Nutmeg~\cite{eastmanNutmegSPICEModels2024}. Additionally, we consider the NNPOps optimizations on CUDA GPUs for the ANI2x model~\cite{galvelisNNPMMAccelerating2023}. All simulations were run in the NVE ensemble at a time step of 1 fs on an NVIDIA GeForce RTX 3070 GPU. Numerical values for all benchmarks are reported in the SI, Tab.~\ref{tab:perf_ms_step} and \ref{tab:perf_ns_day}. 
Up to system sizes of 30 atoms, all models run at a performance of not more than 12 ms/step (8 ns/day at as 1 fs time step). While significantly slower than classical force fields, this performance still marks a speedup of 4 to 5 orders of magnitude compared to QM approaches, where a state-of-the-art approaches for \textit{ab initio} MD simulations of similar size report performances on the order of 100 s/step~\cite{yokelsonPerformanceAnalysisCP2K2022}.

We observe that the performance curves of all models are essentially flat for smaller systems, before exhibiting the expected linear scaling with increasing system size. This plateau is likely due to the combined overhead of repeated CUDA kernel launches and LibTorch API calls, which becomes negligible for larger systems as they transition into a compute-bound regime. This is further supported by the observation that for small NNP region sizes, the NNPOps model is the most performant, almost by a factor of 10. The primary speedup in this implementation is achieved through specialized, fused CUDA operations that efficiently parallelize the 8 neural networks comprising the ensemble ANI2x model while minimizing API interactions. However, the specialized data structures required for this low-latency optimization scale poorly with system size, losing its performance advantage and quickly reaching memory limits. On the other hand, this also indicates that there are potential performance gains to be had in all models by raising (lowering) the small-system plateau, particularly for ML/MM simulation settings. This could be achieved by utilizing modern optimizations like CUDA graphs, and PyTorch's \texttt{compile} and \texttt{AOTInductor} mechanisms that produce optimized GPU kernels, requiring changes to model architecture. 

\section{Conclusion}
In this work, we have introduced a flexible and efficient interface for incorporating neural network potentials into molecular dynamics simulations with GROMACS. By enabling PyTorch-trained models to directly contribute to the force calculations of a subset of atoms or entire systems, this development expands the computational and methodological possibilities for hybrid ML/MM simulations. We have demonstrated that the interface integrates seamlessly with existing GROMACS workflows in a variety of applications, including enhanced sampling techniques alchemical free energy methods and protein--ligand simulations. 

Our results indicate that neural network potentials in our interface are able to reproduce established reference data and, in some cases, improve upon classical force field descriptions, especially in systems where intramolecular interactions are challenging to capture with traditional parameter sets. At the same time, the presented applications also highlight the limitations of the mechanical embedding scheme and underline the need for further development of electrostatic embedding and free energy schemes for NNPs to extend applicability to a wider range of systems~\cite{zinovjevEmleengineFlexibleElectrostatic2024, thurlemannAMPBMSMMMultiscale, harrymooreComputingSolvationFree2026}.

While the computational performance of neural network potentials cannot currently rival that of established and well-validated classical force fields, it is important to highlight their orders-of-magnitude advantage over the \textit{ab initio} QM methods they were trained on. It is also for this reason we believe that in biomolecular contexts, the most promising application of NNPs lies in hybrid ML/MM settings. For these smaller ML system sizes, we observed that nearly all current NNP architectures hit a "speed of light"-like barrier, as computational performance becomes bound by the overhead of LibTorch operation dispatch and kernel launches. Thus, an auspicious avenue for future work lies in the development of direct C++ implementations of specific NNP architectures optimized for inference.

Continued development efforts will also focus on improving the treatment of ML–MM boundaries and long-range electrostatic interactions, improvements regarding computational performance and scaling, and ensuring continued compatibility the rapidly advancing landscape of NNP architectures. Nevertheless, we anticipate that this tool in its current state will accelerate the adoption of ML-based potentials in routine molecular simulations, and create new opportunities for the accurate and efficient modeling of complex biochemical systems at scale. 

\paragraph{Software Availability.}
Capabilities needed to perform all simulations in this paper are available in the recent 2026 version of GROMACS. Extensive examples and guides on how to export NNPs for use with our interface are available as a GitHub repository at \url{https://github.com/lmuellender/gmx-nnpot-tools}, with a persistent record of the version used in this work uploaded to Zenodo at \url{https://doi.org/10.5281/zenodo.19663881}.

\paragraph{Acknowledgments.}
This research was partly supported by the AQTIVATE project (EU Grant Agreement No. 101072344), the Swedish e-Science Research Centre, Vetenskapsrådet (2025-06231), and the BioExcel-3 Center of Excellence (101093290). Computational resources were provided under NAISS project 2025/3-71.  


\printbibliography[heading=bibintoc, title={References}]

@article{abrahamGROMACSHighPerformance2015,
  title = {{{GROMACS}}: {{High}} Performance Molecular Simulations through Multi-Level Parallelism from Laptops to Supercomputers},
  shorttitle = {{{GROMACS}}},
  author = {Abraham, Mark James and Murtola, Teemu and Schulz, Roland and P{\'a}ll, Szil{\'a}rd and Smith, Jeremy C. and Hess, Berk and Lindahl, Erik},
  year = 2015,
  month = sep,
  journal = {SoftwareX},
  volume = {1--2},
  pages = {19--25},
  doi = {10.1016/j.softx.2015.06.001}
}

@article{anstineAIMNet2NeuralNetwork2025,
  title = {{{AIMNet2}}: A Neural Network Potential to Meet Your Neutral, Charged, Organic, and Elemental-Organic Needs},
  shorttitle = {{{AIMNet2}}},
  author = {Anstine, Dylan M. and Zubatyuk, Roman and Isayev, Olexandr},
  year = 2025,
  month = jun,
  journal = {Chemical Science},
  volume = {16},
  number = {23},
  pages = {10228--10244},
  doi = {10.1039/D4SC08572H}
}

@misc{batatiaMACEHigherOrder2023,
  title = {{{MACE}}: {{Higher Order Equivariant Message Passing Neural Networks}} for {{Fast}} and {{Accurate Force Fields}}},
  shorttitle = {{{MACE}}},
  author = {Batatia, Ilyes and Kov{\'a}cs, D{\'a}vid P{\'e}ter and Simm, Gregor N. C. and Ortner, Christoph and Cs{\'a}nyi, G{\'a}bor},
  year = 2023,
  month = jan,
  number = {arXiv:2206.07697},
  eprint = {2206.07697},
  primaryclass = {stat},
  doi = {10.48550/arXiv.2206.07697},
  archiveprefix = {arXiv}
}

@article{batznerE3equivariantGraphNeural2022,
  title = {E(3)-Equivariant Graph Neural Networks for Data-Efficient and Accurate Interatomic Potentials},
  author = {Batzner, Simon and Musaelian, Albert and Sun, Lixin and Geiger, Mario and Mailoa, Jonathan P. and Kornbluth, Mordechai and Molinari, Nicola and Smidt, Tess E. and Kozinsky, Boris},
  year = 2022,
  month = may,
  journal = {Nature Communications},
  volume = {13},
  number = {1},
  pages = {2453},
  doi = {10.1038/s41467-022-29939-5}
}

@article{behlerFourGenerationsHighDimensional2021,
  title = {Four {{Generations}} of {{High-Dimensional Neural Network Potentials}}},
  author = {Behler, J{\"o}rg},
  year = 2021,
  month = aug,
  journal = {Chemical Reviews},
  volume = {121},
  number = {16},
  pages = {10037--10072},
  doi = {10.1021/acs.chemrev.0c00868}
}

@article{behlerGeneralizedNeuralNetworkRepresentation2007,
  title = {Generalized {{Neural-Network Representation}} of {{High-Dimensional Potential-Energy Surfaces}}},
  author = {Behler, J{\"o}rg and Parrinello, Michele},
  year = 2007,
  month = apr,
  journal = {Physical Review Letters},
  volume = {98},
  number = {14},
  pages = {146401},
  doi = {10.1103/PhysRevLett.98.146401}
}

@article{bennettEfficientEstimationFree1976,
  title = {Efficient Estimation of Free Energy Differences from {{Monte Carlo}} Data},
  author = {Bennett, Charles H},
  year = 1976,
  month = oct,
  journal = {Journal of Computational Physics},
  volume = {22},
  number = {2},
  pages = {245--268},
  doi = {10.1016/0021-9991(76)90078-4}
}

@article{bernettiPressureControlUsing2020,
  title = {Pressure Control Using Stochastic Cell Rescaling},
  author = {Bernetti, Mattia and Bussi, Giovanni},
  year = 2020,
  month = sep,
  journal = {The Journal of Chemical Physics},
  volume = {153},
  number = {11},
  pages = {114107},
  doi = {10.1063/5.0020514}
}

@article{bestOptimizationAdditiveCHARMM2012,
  title = {Optimization of the {{Additive CHARMM All-Atom Protein Force Field Targeting Improved Sampling}} of the {{Backbone}} {$\phi$}, {$\psi$} and {{Side-Chain}} {$\chi$}1 and {$\chi$}2 {{Dihedral Angles}}},
  author = {Best, Robert B. and Zhu, Xiao and Shim, Jihyun and Lopes, Pedro E. M. and Mittal, Jeetain and Feig, Michael and MacKerell, Alexander D. Jr.},
  year = 2012,
  month = sep,
  journal = {Journal of Chemical Theory and Computation},
  volume = {8},
  number = {9},
  pages = {3257--3273},
  doi = {10.1021/ct300400x}
}

@article{bouyssetProLIFLibraryEncode2021,
  title = {{{ProLIF}}: A Library to Encode Molecular Interactions as Fingerprints},
  shorttitle = {{{ProLIF}}},
  author = {Bouysset, C{\'e}dric and Fiorucci, S{\'e}bastien},
  year = 2021,
  month = sep,
  journal = {Journal of Cheminformatics},
  volume = {13},
  number = {1},
  pages = {72},
  doi = {10.1186/s13321-021-00548-6}
}

@article{boycePredictingLigandBinding2009,
  title = {Predicting {{Ligand Binding Affinity}} with {{Alchemical Free Energy Methods}} in a {{Polar Model Binding Site}}},
  author = {Boyce, Sarah E. and Mobley, David L. and Rocklin, Gabriel J. and Graves, Alan P. and Dill, Ken A. and Shoichet, Brian K.},
  year = 2009,
  month = dec,
  journal = {Journal of Molecular Biology},
  volume = {394},
  number = {4},
  pages = {747--763},
  doi = {10.1016/j.jmb.2009.09.049}
}

@article{bussiCanonicalSamplingVelocity2007,
  title = {Canonical Sampling through Velocity Rescaling},
  author = {Bussi, Giovanni and Donadio, Davide and Parrinello, Michele},
  year = 2007,
  month = jan,
  journal = {The Journal of Chemical Physics},
  volume = {126},
  number = {1},
  pages = {014101},
  doi = {10.1063/1.2408420}
}

@article{chmielaMachineLearningAccurate2017,
  title = {Machine Learning of Accurate Energy-Conserving Molecular Force Fields},
  author = {Chmiela, Stefan and Tkatchenko, Alexandre and Sauceda, Huziel E. and Poltavsky, Igor and Sch{\"u}tt, Kristof T. and M{\"u}ller, Klaus-Robert},
  year = 2017,
  month = may,
  journal = {Science Advances},
  volume = {3},
  number = {5},
  pages = {e1603015},
  doi = {10.1126/sciadv.1603015}
}

@article{devereuxExtendingApplicabilityANI2020,
  title = {Extending the {{Applicability}} of the {{ANI Deep Learning Molecular Potential}} to {{Sulfur}} and {{Halogens}}},
  author = {Devereux, Christian and Smith, Justin S. and Huddleston, Kate K. and Barros, Kipton and Zubatyuk, Roman and Isayev, Olexandr and Roitberg, Adrian E.},
  year = 2020,
  month = jul,
  journal = {Journal of Chemical Theory and Computation},
  volume = {16},
  number = {7},
  pages = {4192--4202},
  doi = {10.1021/acs.jctc.0c00121}
}

@article{eastmanNutmegSPICEModels2024,
  title = {Nutmeg and {{SPICE}}: {{Models}} and {{Data}} for {{Biomolecular Machine Learning}}},
  shorttitle = {Nutmeg and {{SPICE}}},
  author = {Eastman, Peter and Pritchard, Benjamin P. and Chodera, John D. and Markland, Thomas E.},
  year = 2024,
  month = oct,
  journal = {Journal of Chemical Theory and Computation},
  volume = {20},
  number = {19},
  pages = {8583--8593},
  doi = {10.1021/acs.jctc.4c00794}
}

@article{eastmanOpenMM8Molecular2024,
  title = {{{OpenMM}} 8: {{Molecular Dynamics Simulation}} with {{Machine Learning Potentials}}},
  shorttitle = {{{OpenMM}} 8},
  author = {Eastman, Peter and Galvelis, Raimondas and Pel{\'a}ez, Ra{\'u}l P. and Abreu, Charlles R. A. and Farr, Stephen E. and Gallicchio, Emilio and Gorenko, Anton and Henry, Michael M. and Hu, Frank and Huang, Jing and Kr{\"a}mer, Andreas and Michel, Julien and Mitchell, Joshua A. and Pande, Vijay S. and Rodrigues, Jo{\~a}o PGLM and {Rodriguez-Guerra}, Jaime and Simmonett, Andrew C. and Singh, Sukrit and Swails, Jason and Turner, Philip and Wang, Yuanqing and Zhang, Ivy and Chodera, John D. and De Fabritiis, Gianni and Markland, Thomas E.},
  year = 2024,
  month = jan,
  journal = {The Journal of Physical Chemistry B},
  volume = {128},
  number = {1},
  pages = {109--116},
  doi = {10.1021/acs.jpcb.3c06662}
}

@article{fieldDynamoLibraryMolecular2000,
  title = {The Dynamo Library for Molecular Simulations Using Hybrid Quantum Mechanical and Molecular Mechanical Potentials},
  author = {Field, Martin J. and Albe, Marc and Bret, C{\'e}line and {Proust-De Martin}, Flavien and Thomas, Aline},
  year = 2000,
  journal = {Journal of Computational Chemistry},
  volume = {21},
  number = {12},
  pages = {1088--1100},
  doi = {10.1002/1096-987X(200009)21:12<1088::AID-JCC5>3.0.CO;2-8}
}

@article{fuchsChemtraindeployParallelScalable2025,
  title = {Chemtrain-Deploy: {{A Parallel}} and {{Scalable Framework}} for {{Machine Learning Potentials}} in {{Million-Atom MD Simulations}}},
  shorttitle = {Chemtrain-Deploy},
  author = {Fuchs, Paul and Chen, Weilong and Thaler, Stephan and Zavadlav, Julija},
  year = 2025,
  month = aug,
  journal = {Journal of Chemical Theory and Computation},
  volume = {21},
  number = {15},
  pages = {7550--7560},
  doi = {10.1021/acs.jctc.5c00996}
}

@article{galvelisNNPMMAccelerating2023,
  title = {{{NNP}}/{{MM}}: {{Accelerating Molecular Dynamics Simulations}} with {{Machine Learning Potentials}} and {{Molecular Mechanics}}},
  shorttitle = {{{NNP}}/{{MM}}},
  author = {Galvelis, Raimondas and {Varela-Rial}, Alejandro and Doerr, Stefan and Fino, Roberto and Eastman, Peter and Markland, Thomas E. and Chodera, John D. and De Fabritiis, Gianni},
  year = 2023,
  month = sep,
  journal = {Journal of Chemical Information and Modeling},
  volume = {63},
  number = {18},
  pages = {5701--5708},
  doi = {10.1021/acs.jcim.3c00773}
}

@article{gasteggerMachineLearningSolvent2021,
  title = {Machine Learning of Solvent Effects on Molecular Spectra and Reactions},
  author = {Gastegger, Michael and T.~Sch{\"u}tt, Kristof and M{\"u}ller, Klaus-Robert},
  year = 2021,
  journal = {Chemical Science},
  volume = {12},
  number = {34},
  pages = {11473--11483},
  doi = {10.1039/D1SC02742E}
}

@article{grassanoAssessmentEmbeddingSchemes2024,
  title = {Assessment of {{Embedding Schemes}} in a {{Hybrid Machine Learning}}/{{Classical Potentials}} ({{ML}}/{{MM}}) {{Approach}}},
  author = {Grassano, Juan S. and Pickering, Ignacio and Roitberg, Adrian E. and Gonz{\'a}lez Lebrero, Mariano C. and Estrin, Dario A. and Semelak, Jonathan A.},
  year = 2024,
  month = may,
  journal = {Journal of Chemical Information and Modeling},
  volume = {64},
  number = {10},
  pages = {4047--4058},
  doi = {10.1021/acs.jcim.4c00478}
}

@article{haghiriANIEFPModeling2024,
  title = {{{ANI}}/{{EFP}}: {{Modeling Long-Range Interactions}} in {{ANI Neural Network}} with {{Effective Fragment Potentials}}},
  shorttitle = {{{ANI}}/{{EFP}}},
  author = {Haghiri, Shahed and Viquez Rojas, Claudia and Bhat, Sriram and Isayev, Olexandr and Slipchenko, Lyudmila},
  year = 2024,
  month = oct,
  journal = {Journal of Chemical Theory and Computation},
  volume = {20},
  number = {20},
  pages = {9138--9147},
  doi = {10.1021/acs.jctc.4c01052}
}

@article{harrymooreComputingSolvationFree2026,
  title = {Computing {{Solvation Free Energies}} of {{Small Molecules}} with {{Experimental Accuracy}}},
  author = {Harry Moore, J. and Cole, Daniel J. and Cs{\'a}nyi, G{\'a}bor},
  year = 2026,
  month = jan,
  journal = {Journal of the American Chemical Society},
  doi = {10.1021/jacs.5c10940}
}

@article{hjorthlarsenAtomicSimulationEnvironment2017,
  title = {The Atomic Simulation Environment---a {{Python}} Library for Working with Atoms},
  author = {Hjorth Larsen, Ask and J{\o}rgen Mortensen, Jens and Blomqvist, Jakob and Castelli, Ivano E and Christensen, Rune and Du{\l}ak, Marcin and Friis, Jesper and Groves, Michael N and Hammer, Bj{\o}rk and Hargus, Cory and Hermes, Eric D and Jennings, Paul C and Bjerre Jensen, Peter and Kermode, James and Kitchin, John R and Leonhard Kolsbjerg, Esben and Kubal, Joseph and Kaasbjerg, Kristen and Lysgaard, Steen and Bergmann Maronsson, J{\'o}n and Maxson, Tristan and Olsen, Thomas and Pastewka, Lars and Peterson, Andrew and Rostgaard, Carsten and Schi{\o}tz, Jakob and Sch{\"u}tt, Ole and Strange, Mikkel and Thygesen, Kristian S and Vegge, Tejs and Vilhelmsen, Lasse and Walter, Michael and Zeng, Zhenhua and Jacobsen, Karsten W},
  year = 2017,
  month = jun,
  journal = {Journal of Physics: Condensed Matter},
  volume = {29},
  number = {27},
  pages = {273002},
  doi = {10.1088/1361-648X/aa680e}
}

@misc{jax2018github,
  title = {{{JAX}}: Composable Transformations of {{Python}}+{{NumPy}} Programs},
  author = {Bradbury, James and Frostig, Roy and Hawkins, Peter and Johnson, Matthew James and Leary, Chris and Maclaurin, Dougal and Necula, George and Paszke, Adam and VanderPlas, Jake and {Wanderman-Milne}, Skye and Zhang, Qiao},
  year = 2018
}

@article{jorgensenComparisonSimplePotential1983,
  title = {Comparison of Simple Potential Functions for Simulating Liquid Water},
  author = {Jorgensen, William L. and Chandrasekhar, Jayaraman and Madura, Jeffry D. and Impey, Roger W. and Klein, Michael L.},
  year = 1983,
  month = jul,
  journal = {The Journal of Chemical Physics},
  volume = {79},
  number = {2},
  pages = {926--935},
  doi = {10.1063/1.445869}
}

@article{kabyldaMolecularSimulationsPretrained2025,
  title = {Molecular {{Simulations}} with a {{Pretrained Neural Network}} and {{Universal Pairwise Force Fields}}},
  author = {Kabylda, Adil and Frank, J. Thorben and {Su{\'a}rez-Dou}, Sergio and Khabibrakhmanov, Almaz and Medrano Sandonas, Leonardo and Unke, Oliver T. and Chmiela, Stefan and M{\"u}ller, Klaus-Robert and Tkatchenko, Alexandre},
  year = 2025,
  month = sep,
  journal = {Journal of the American Chemical Society},
  volume = {147},
  number = {37},
  pages = {33723--33734},
  doi = {10.1021/jacs.5c09558}
}

@article{karwounopoulosInsightsChallengesCorrecting2024,
  title = {Insights and {{Challenges}} in {{Correcting Force Field Based Solvation Free Energies Using}} a {{Neural Network Potential}}},
  author = {Karwounopoulos, Johannes and Wu, Zhiyi and Tkaczyk, Sara and Wang, Shuzhe and Baskerville, Adam and Ranasinghe, Kavindri and Langer, Thierry and Wood, Geoffrey P. F. and Wieder, Marcus and Boresch, Stefan},
  year = 2024,
  month = jul,
  journal = {The Journal of Physical Chemistry B},
  volume = {128},
  number = {28},
  pages = {6693--6703},
  doi = {10.1021/acs.jpcb.4c01417}
}

@article{koFourthgenerationHighdimensionalNeural2021,
  title = {A Fourth-Generation High-Dimensional Neural Network Potential with Accurate Electrostatics Including Non-Local Charge Transfer},
  author = {Ko, Tsz Wai and Finkler, Jonas A. and Goedecker, Stefan and Behler, J{\"o}rg},
  year = 2021,
  month = jan,
  journal = {Nature Communications},
  volume = {12},
  number = {1},
  pages = {398},
  doi = {10.1038/s41467-020-20427-2}
}

@article{kovacsMACEOFFShortRangeTransferable2025,
  title = {{{MACE-OFF}}: {{Short-Range Transferable Machine Learning Force Fields}} for {{Organic Molecules}}},
  shorttitle = {{{MACE-OFF}}},
  author = {Kov{\'a}cs, D{\'a}vid P{\'e}ter and Moore, J. Harry and Browning, Nicholas J. and Batatia, Ilyes and Horton, Joshua T. and Pu, Yixuan and Kapil, Venkat and Witt, William C. and Magd{\u a}u, Ioan-Bogdan and Cole, Daniel J. and Cs{\'a}nyi, G{\'a}bor},
  year = 2025,
  month = may,
  journal = {Journal of the American Chemical Society},
  volume = {147},
  number = {21},
  pages = {17598--17611},
  doi = {10.1021/jacs.4c07099}
}

@article{kulikHowLargeShould2016,
  title = {How {{Large Should}} the {{QM Region Be}} in {{QM}}/{{MM Calculations}}? {{The Case}} of {{Catechol O-Methyltransferase}}},
  shorttitle = {How {{Large Should}} the {{QM Region Be}} in {{QM}}/{{MM Calculations}}?},
  author = {Kulik, Heather J. and Zhang, Jianyu and Klinman, Judith P. and Mart{\'i}nez, Todd J.},
  year = 2016,
  month = nov,
  journal = {The Journal of Physical Chemistry B},
  volume = {120},
  number = {44},
  pages = {11381--11394},
  doi = {10.1021/acs.jpcb.6b07814}
}

@article{laheySimulatingProteinLigand2020,
  title = {Simulating Protein--Ligand Binding with Neural Network Potentials},
  author = {Lahey, Shae-Lynn J. and Rowley, Christopher N.},
  year = 2020,
  month = mar,
  journal = {Chemical Science},
  volume = {11},
  number = {9},
  pages = {2362--2368},
  doi = {10.1039/C9SC06017K}
}

@article{lierBuRNNBufferRegion2022,
  title = {{{BuRNN}}: {{Buffer Region Neural Network Approach}} for {{Polarizable-Embedding Neural Network}}/{{Molecular Mechanics Simulations}}},
  shorttitle = {{{BuRNN}}},
  author = {Lier, Bettina and Poliak, Peter and Marquetand, Philipp and Westermayr, Julia and Oostenbrink, Chris},
  year = 2022,
  month = may,
  journal = {The Journal of Physical Chemistry Letters},
  volume = {13},
  number = {17},
  pages = {3812--3818},
  doi = {10.1021/acs.jpclett.2c00654}
}

@article{lindahlAcceleratedWeightHistogram2014,
  title = {Accelerated Weight Histogram Method for Exploring Free Energy Landscapes},
  author = {Lindahl, V. and Lidmar, J. and Hess, B.},
  year = 2014,
  month = jul,
  journal = {The Journal of Chemical Physics},
  volume = {141},
  number = {4},
  pages = {044110},
  doi = {10.1063/1.4890371}
}

@article{linQMMMWhat2007,
  title = {{{QM}}/{{MM}}: What Have We Learned, Where Are We, and Where Do We Go from Here?},
  shorttitle = {{{QM}}/{{MM}}},
  author = {Lin, Hai and Truhlar, Donald G.},
  year = 2007,
  month = feb,
  journal = {Theoretical Chemistry Accounts},
  volume = {117},
  number = {2},
  pages = {185--199},
  doi = {10.1007/s00214-006-0143-z}
}

@article{lundborgSkinPermeabilityPrediction2022,
  title = {Skin Permeability Prediction with {{MD}} Simulation Sampling Spatial and Alchemical Reaction Coordinates},
  author = {Lundborg, Magnus and Wennberg, Christian and Lidmar, Jack and Hess, Berk and Lindahl, Erik and Norl{\'e}n, Lars},
  year = 2022,
  month = oct,
  journal = {Biophysical Journal},
  volume = {121},
  number = {20},
  pages = {3837--3849},
  doi = {10.1016/j.bpj.2022.09.009},
  pmid = {36104960}
}

@article{madisonSolventdependentConformationalDistributions1980,
  title = {Solvent-Dependent Conformational Distributions of Some Dipeptides},
  author = {Madison, Vincent and Kopple, Kenneth D.},
  year = 1980,
  month = jul,
  journal = {Journal of the American Chemical Society},
  volume = {102},
  number = {15},
  pages = {4855--4863},
  doi = {10.1021/ja00535a001}
}

@article{mishinMachinelearningInteratomicPotentials2021,
  title = {Machine-Learning Interatomic Potentials for Materials Science},
  author = {Mishin, Y.},
  year = 2021,
  month = aug,
  journal = {Acta Materialia},
  volume = {214},
  pages = {116980},
  doi = {10.1016/j.actamat.2021.116980}
}

@article{mobleyFreeSolvDatabaseExperimental2014,
  title = {{{FreeSolv}}: A Database of Experimental and Calculated Hydration Free Energies, with Input Files},
  shorttitle = {{{FreeSolv}}},
  author = {Mobley, David L. and Guthrie, J. Peter},
  year = 2014,
  month = jul,
  journal = {Journal of Computer-Aided Molecular Design},
  volume = {28},
  number = {7},
  pages = {711--720},
  doi = {10.1007/s10822-014-9747-x}
}

@article{musaelianLearningLocalEquivariant2023,
  title = {Learning Local Equivariant Representations for Large-Scale Atomistic Dynamics},
  author = {Musaelian, Albert and Batzner, Simon and Johansson, Anders and Sun, Lixin and Owen, Cameron J. and Kornbluth, Mordechai and Kozinsky, Boris},
  year = 2023,
  month = feb,
  journal = {Nature Communications},
  volume = {14},
  number = {1},
  pages = {579},
  doi = {10.1038/s41467-023-36329-y}
}

@incollection{paszkePyTorchImperativeStyle2019,
  title = {{{PyTorch}}: An Imperative Style, High-Performance Deep Learning Library},
  shorttitle = {{{PyTorch}}},
  booktitle = {Proceedings of the 33rd {{International Conference}} on {{Neural Information Processing Systems}}},
  author = {Paszke, Adam and Gross, Sam and Massa, Francisco and Lerer, Adam and Bradbury, James and Chanan, Gregory and Killeen, Trevor and Lin, Zeming and Gimelshein, Natalia and Antiga, Luca and Desmaison, Alban and K{\"o}pf, Andreas and Yang, Edward and DeVito, Zach and Raison, Martin and Tejani, Alykhan and Chilamkurthy, Sasank and Steiner, Benoit and Fang, Lu and Bai, Junjie and Chintala, Soumith},
  year = 2019,
  month = dec,
  number = {721},
  pages = {8026--8037}
}

@article{pickeringTorchANI20Extensible2025,
  title = {{{TorchANI}} 2.0: {{An Extensible}}, {{High-Performance Library}} for the {{Design}}, {{Training}}, and {{Use}} of {{NN-IPs}}},
  shorttitle = {{{TorchANI}} 2.0},
  author = {Pickering, Ignacio and Xue, Jinze and Huddleston, Kate and Terrel, Nicholas and Roitberg, Adrian E.},
  year = 2025,
  month = nov,
  journal = {Journal of Chemical Information and Modeling},
  volume = {65},
  number = {21},
  pages = {11656--11671},
  doi = {10.1021/acs.jcim.5c01853}
}

@article{pickeringTorchANIAmberBridgingNeural2025,
  title = {{{TorchANI-Amber}}: {{Bridging Neural Network Potentials}} and {{Classical Biomolecular Simulations}}},
  shorttitle = {{{TorchANI-Amber}}},
  author = {Pickering, Ignacio and Semelak, Jonathan A. and Xue, Jinze and Roitberg, Adrian E.},
  year = 2025,
  month = nov,
  journal = {The Journal of Physical Chemistry B},
  volume = {129},
  number = {46},
  pages = {11927--11938},
  doi = {10.1021/acs.jpcb.5c05725}
}

@article{pultarNeuralNetworkPotential2025,
  title = {Neural {{Network Potential}} with {{Multiresolution Approach Enables Accurate Prediction}} of {{Reaction Free Energies}} in {{Solution}}},
  author = {Pultar, Felix and Th{\"u}rlemann, Moritz and Gordiy, Igor and Doloszeski, Eva and Riniker, Sereina},
  year = 2025,
  month = feb,
  journal = {Journal of the American Chemical Society},
  volume = {147},
  number = {8},
  pages = {6835--6856},
  doi = {10.1021/jacs.4c17015},
  pmcid = {PMC11869291},
  pmid = {39961342}
}

@misc{ReferenceManualGROMACS,
  title = {Reference Manual - GROMACS 2026.0 Documentation},
  url = {https://manual.gromacs.org/documentation/current/reference-manual/index.html}
}

@misc{rufaChemicalAccuracyAlchemical2020,
  title = {Towards Chemical Accuracy for Alchemical Free Energy Calculations with Hybrid Physics-Based Machine Learning / Molecular Mechanics Potentials},
  author = {Rufa, Dominic A. and Macdonald, Hannah E. Bruce and Fass, Josh and Wieder, Marcus and Grinaway, Patrick B. and Roitberg, Adrian E. and Isayev, Olexandr and Chodera, John D.},
  year = 2020,
  month = jul,
  number = {2020.07.29.227959},
  eprint = {2020.07.29.227959},
  doi = {10.1101/2020.07.29.227959},
  archiveprefix = {bioRxiv}
}

@article{sabaneszariquieyEnhancingProteinLigand2024,
  title = {Enhancing {{Protein}}--{{Ligand Binding Affinity Predictions Using Neural Network Potentials}}},
  author = {Saban{\'e}s Zariquiey, Francesc and Galvelis, Raimondas and Gallicchio, Emilio and Chodera, John D. and Markland, Thomas E. and De Fabritiis, Gianni},
  year = 2024,
  month = mar,
  journal = {Journal of Chemical Information and Modeling},
  volume = {64},
  number = {5},
  pages = {1481--1485},
  doi = {10.1021/acs.jcim.3c02031}
}

@article{semelakAdvancingMultiscaleMolecular2025,
  title = {Advancing {{Multiscale Molecular Modeling}} with {{Machine Learning-Derived Electrostatics}}},
  author = {Semelak, Jonathan A. and Pickering, Ignacio and Huddleston, Kate and Olmos, Justo and Grassano, Juan Santiago and Clemente, Camila Mara and Drusin, Salvador I. and Marti, Marcelo and Gonzalez Lebrero, Mariano Camilo and Roitberg, Adrian E. and Estrin, Dario A.},
  year = 2025,
  month = mar,
  journal = {Journal of Chemical Theory and Computation},
  doi = {10.1021/acs.jctc.4c01792}
}

@article{sennQMMMMethods2009,
  title = {{{QM}}/{{MM Methods}} for {{Biomolecular Systems}}},
  author = {Senn, Hans Martin and Thiel, Walter},
  year = 2009,
  journal = {Angewandte Chemie International Edition},
  volume = {48},
  number = {7},
  pages = {1198--1229},
  doi = {10.1002/anie.200802019}
}

@article{shaiduIncorporatingLongrangeElectrostatics2024,
  title = {Incorporating Long-Range Electrostatics in Neural Network Potentials via Variational Charge Equilibration from Shortsighted Ingredients},
  author = {Shaidu, Yusuf and Pellegrini, Franco and K{\"u}{\c c}{\"u}kbenli, Emine and Lot, Ruggero and {de Gironcoli}, Stefano},
  year = 2024,
  month = mar,
  journal = {npj Computational Materials},
  volume = {10},
  number = {1},
  pages = {47},
  doi = {10.1038/s41524-024-01225-6}
}

@article{singhCombinedInitioQuantum1986,
  title = {A Combined Ab Initio Quantum Mechanical and Molecular Mechanical Method for Carrying out Simulations on Complex Molecular Systems: {{Applications}} to the {{CH3Cl}} + {{Cl}}- Exchange Reaction and Gas Phase Protonation of Polyethers},
  shorttitle = {A Combined Ab Initio Quantum Mechanical and Molecular Mechanical Method for Carrying out Simulations on Complex Molecular Systems},
  author = {Singh, U. Chandra and Kollman, Peter A.},
  year = 1986,
  journal = {Journal of Computational Chemistry},
  volume = {7},
  number = {6},
  pages = {718--730},
  doi = {10.1002/jcc.540070604}
}

@article{smithANI1ExtensibleNeural2017,
  title = {{{ANI-1}}: An Extensible Neural Network Potential with {{DFT}} Accuracy at Force Field Computational Cost},
  shorttitle = {{{ANI-1}}},
  author = {Smith, J. S. and Isayev, O. and Roitberg, A. E.},
  year = 2017,
  month = mar,
  journal = {Chemical Science},
  volume = {8},
  number = {4},
  pages = {3192--3203},
  doi = {10.1039/C6SC05720A}
}

@article{thompsonLAMMPSFlexibleSimulation2022,
  title = {{{LAMMPS}} - a Flexible Simulation Tool for Particle-Based Materials Modeling at the Atomic, Meso, and Continuum Scales},
  author = {Thompson, Aidan P. and Aktulga, H. Metin and Berger, Richard and Bolintineanu, Dan S. and Brown, W. Michael and Crozier, Paul S. and {in 't Veld}, Pieter J. and Kohlmeyer, Axel and Moore, Stan G. and Nguyen, Trung Dac and Shan, Ray and Stevens, Mark J. and Tranchida, Julien and Trott, Christian and Plimpton, Steven J.},
  year = 2022,
  month = feb,
  journal = {Computer Physics Communications},
  volume = {271},
  pages = {108171},
  doi = {10.1016/j.cpc.2021.108171}
}

@misc{thurlemannAMPBMSMMMultiscale,
  title = {{{AMP-BMS}}/{{MM}}: {{A Multiscale Neural Network Potential}} for the {{Fast}} and {{Accurate Simulation}} of {{Protein Dynamics}} and {{Enzymatic Reactions}}},
  shorttitle = {{{AMP-BMS}}/{{MM}}},
  author = {Th{\"u}rlemann, Moritz and Pultar, Felix and Gordiy, Igor and Ruijsenaars, Enrico and Riniker, Sereina},
  year = 2026,
  number = {10.26434/chemrxiv-2026-kx9w0},
  eprint = {10.26434/chemrxiv-2026-kx9w0},
  doi = {10.26434/chemrxiv-2026-kx9w0},
  archiveprefix = {ChemRxiv}
}

@article{tianFf19SBAminoAcidSpecificProtein2020,
  title = {{{ff19SB}}: {{Amino-Acid-Specific Protein Backbone Parameters Trained}} against {{Quantum Mechanics Energy Surfaces}} in {{Solution}}},
  shorttitle = {{{ff19SB}}},
  author = {Tian, Chuan and Kasavajhala, Koushik and Belfon, Kellon A. A. and Raguette, Lauren and Huang, He and Migues, Angela N. and Bickel, John and Wang, Yuzhang and Pincay, Jorge and Wu, Qin and Simmerling, Carlos},
  year = 2020,
  month = jan,
  journal = {Journal of Chemical Theory and Computation},
  volume = {16},
  number = {1},
  pages = {528--552},
  doi = {10.1021/acs.jctc.9b00591}
}

@article{tobiasConformationalEquilibriumAlanine1992,
  title = {Conformational Equilibrium in the Alanine Dipeptide in the Gas Phase and Aqueous Solution: A Comparison of Theoretical Results},
  shorttitle = {Conformational Equilibrium in the Alanine Dipeptide in the Gas Phase and Aqueous Solution},
  author = {Tobias, Douglas J. and Brooks, Charles L. III},
  year = 1992,
  month = apr,
  journal = {The Journal of Physical Chemistry},
  volume = {96},
  number = {9},
  pages = {3864--3870},
  doi = {10.1021/j100188a054}
}

@article{torchani,
  title = {{{TorchANI}}: A Free and Open Source {{PyTorch-based}} Deep Learning Implementation of the {{ANI}} Neural Network Potentials},
  author = {Gao, Xiang and Ramezanghorbani, Farhad and Isayev, Olexandr and Smith, Justin S. and Roitberg, Adrian E.},
  year = 2020,
  journal = {Journal of Chemical Information and Modeling},
  volume = {60},
  number = {7},
  pages = {3408--3415},
  doi = {10.1021/acs.jcim.0c00451}
}

@article{unkeBiomolecularDynamicsMachinelearned2024,
  title = {Biomolecular Dynamics with Machine-Learned Quantum-Mechanical Force Fields Trained on Diverse Chemical Fragments},
  author = {Unke, Oliver T. and St{\"o}hr, Martin and Ganscha, Stefan and Unterthiner, Thomas and Maennel, Hartmut and Kashubin, Sergii and Ahlin, Daniel and Gastegger, Michael and Medrano Sandonas, Leonardo and Berryman, Joshua T. and Tkatchenko, Alexandre and M{\"u}ller, Klaus-Robert},
  year = 2024,
  month = apr,
  journal = {Science Advances},
  volume = {10},
  number = {14},
  pages = {eadn4397},
  doi = {10.1126/sciadv.adn4397}
}

@article{unkeMachineLearningForce2021,
  title = {Machine {{Learning Force Fields}}},
  author = {Unke, Oliver T. and Chmiela, Stefan and Sauceda, Huziel E. and Gastegger, Michael and Poltavsky, Igor and Sch{\"u}tt, Kristof T. and Tkatchenko, Alexandre and M{\"u}ller, Klaus-Robert},
  year = 2021,
  month = aug,
  journal = {Chemical Reviews},
  volume = {121},
  number = {16},
  pages = {10142--10186},
  doi = {10.1021/acs.chemrev.0c01111}
}

@article{unkePhysNetNeuralNetwork2019,
  title = {{{PhysNet}}: {{A Neural Network}} for {{Predicting Energies}}, {{Forces}}, {{Dipole Moments}}, and {{Partial Charges}}},
  shorttitle = {{{PhysNet}}},
  author = {Unke, Oliver T. and Meuwly, Markus},
  year = 2019,
  month = jun,
  journal = {Journal of Chemical Theory and Computation},
  volume = {15},
  number = {6},
  pages = {3678--3693},
  doi = {10.1021/acs.jctc.9b00181}
}

@article{unkeSpookyNetLearningForce2021,
  title = {{{SpookyNet}}: {{Learning}} Force Fields with Electronic Degrees of Freedom and Nonlocal Effects},
  shorttitle = {{{SpookyNet}}},
  author = {Unke, Oliver T. and Chmiela, Stefan and Gastegger, Michael and Sch{\"u}tt, Kristof T. and Sauceda, Huziel E. and M{\"u}ller, Klaus-Robert},
  year = 2021,
  month = dec,
  journal = {Nature Communications},
  volume = {12},
  number = {1},
  pages = {7273},
  doi = {10.1038/s41467-021-27504-0}
}

@article{vanommeslaegheCHARMMGeneralForce2010,
  title = {{{CHARMM}} General Force Field: {{A}} Force Field for Drug-like Molecules Compatible with the {{CHARMM}} All-Atom Additive Biological Force Fields},
  shorttitle = {{{CHARMM}} General Force Field},
  author = {Vanommeslaeghe, K. and Hatcher, E. and Acharya, C. and Kundu, S. and Zhong, S. and Shim, J. and Darian, E. and Guvench, O. and Lopes, P. and Vorobyov, I. and Mackerell Jr., A. D.},
  year = 2010,
  journal = {Journal of Computational Chemistry},
  volume = {31},
  number = {4},
  pages = {671--690},
  doi = {10.1002/jcc.21367}
}

@article{wangDevelopmentTestingGeneral2004,
  title = {Development and Testing of a General Amber Force Field},
  author = {Wang, Junmei and Wolf, Romain M. and Caldwell, James W. and Kollman, Peter A. and Case, David A.},
  year = 2004,
  journal = {Journal of Computational Chemistry},
  volume = {25},
  number = {9},
  pages = {1157--1174},
  doi = {10.1002/jcc.20035}
}

@article{warshelTheoreticalStudiesEnzymic1976a,
  title = {Theoretical Studies of Enzymic Reactions: {{Dielectric}}, Electrostatic and Steric Stabilization of the Carbonium Ion in the Reaction of Lysozyme},
  shorttitle = {Theoretical Studies of Enzymic Reactions},
  author = {Warshel, A. and Levitt, M.},
  year = 1976,
  month = may,
  journal = {Journal of Molecular Biology},
  volume = {103},
  number = {2},
  pages = {227--249},
  doi = {10.1016/0022-2836(76)90311-9}
}

@article{yaoTensorMol01ModelChemistry2018,
  title = {The {{TensorMol-0}}.1 Model Chemistry: A Neural Network Augmented with Long-Range Physics},
  shorttitle = {The {{TensorMol-0}}.1 Model Chemistry},
  author = {Yao, Kun and Herr, John E. and Toth, David W. and Mckintyre, Ryker and Parkhill, John},
  year = 2018,
  month = feb,
  journal = {Chemical Science},
  volume = {9},
  number = {8},
  pages = {2261--2269},
  doi = {10.1039/C7SC04934J}
}

@article{yokelsonPerformanceAnalysisCP2K2022,
  title = {Performance {{Analysis}} of {{CP2K Code}} for {{Ab Initio Molecular Dynamics}} on {{CPUs}} and {{GPUs}}},
  author = {Yokelson, Dewi and Tkachenko, Nikolay V. and Robey, Robert and Li, Ying Wai and Dub, Pavel A.},
  year = 2022,
  month = may,
  journal = {Journal of Chemical Information and Modeling},
  volume = {62},
  number = {10},
  pages = {2378--2386},
  doi = {10.1021/acs.jcim.1c01538}
}

@article{zengDeePMDkitV3MultipleBackend2025,
  title = {{{DeePMD-kit}} v3: {{A Multiple-Backend Framework}} for {{Machine Learning Potentials}}},
  shorttitle = {{{DeePMD-kit}} V3},
  author = {Zeng, Jinzhe and Zhang, Duo and Peng, Anyang and Zhang, Xiangyu and He, Sensen and Wang, Yan and Liu, Xinzijian and Bi, Hangrui and Li, Yifan and Cai, Chun and Zhang, Chengqian and Du, Yiming and Zhu, Jia-Xin and Mo, Pinghui and Huang, Zhengtao and Zeng, Qiyu and Shi, Shaochen and Qin, Xuejian and Yu, Zhaoxi and Luo, Chenxing and Ding, Ye and Liu, Yun-Pei and Shi, Ruosong and Wang, Zhenyu and Bore, Sigbj{\o}rn L{\o}land and Chang, Junhan and Deng, Zhe and Ding, Zhaohan and Han, Siyuan and Jiang, Wanrun and Ke, Guolin and Liu, Zhaoqing and Lu, Denghui and Muraoka, Koki and Oliaei, Hananeh and Singh, Anurag Kumar and Que, Haohui and Xu, Weihong and Xu, Zhangmancang and Zhuang, Yong-Bin and Dai, Jiayu and Giese, Timothy J. and Jia, Weile and Xu, Ben and York, Darrin M. and Zhang, Linfeng and Wang, Han},
  year = 2025,
  month = may,
  journal = {Journal of Chemical Theory and Computation},
  volume = {21},
  number = {9},
  pages = {4375--4385},
  doi = {10.1021/acs.jctc.5c00340}
}

@article{zinovjevElectrostaticEmbeddingMachine2023,
  title = {Electrostatic {{Embedding}} of {{Machine Learning Potentials}}},
  author = {Zinovjev, Kirill},
  year = 2023,
  month = mar,
  journal = {Journal of Chemical Theory and Computation},
  volume = {19},
  number = {6},
  pages = {1888--1897},
  doi = {10.1021/acs.jctc.2c00914}
}

@article{zinovjevEmleengineFlexibleElectrostatic2024,
  title = {Emle-Engine: {{A Flexible Electrostatic Machine Learning Embedding Package}} for {{Multiscale Molecular Dynamics Simulations}}},
  shorttitle = {Emle-Engine},
  author = {Zinovjev, Kirill and Hedges, Lester and Montagud Andreu, Rub{\'e}n and Woods, Christopher and Tu{\~n}{\'o}n, I{\~n}aki and {van der Kamp}, Marc W.},
  year = 2024,
  month = jun,
  journal = {Journal of Chemical Theory and Computation},
  volume = {20},
  number = {11},
  pages = {4514--4522},
  doi = {10.1021/acs.jctc.4c00248}
}

@article{zinovjevImprovedDescriptionEnvironment2025,
  title = {Improved {{Description}} of {{Environment}} and {{Vibronic Effects}} with {{Electrostatically Embedded ML Potentials}}},
  author = {Zinovjev, Kirill and Curutchet, Carles},
  year = 2025,
  month = jan,
  journal = {The Journal of Physical Chemistry Letters},
  volume = {16},
  number = {3},
  pages = {774--781},
  doi = {10.1021/acs.jpclett.4c02949}
}

@article{zlobinChallengesProteinQM2023,
  title = {Challenges in {{Protein QM}}/{{MM Simulations}} with {{Intra-Backbone Link Atoms}}},
  author = {Zlobin, Alexander and Belyaeva, Julia and Golovin, Andrey},
  year = 2023,
  month = jan,
  journal = {Journal of Chemical Information and Modeling},
  volume = {63},
  number = {2},
  pages = {546--560},
  doi = {10.1021/acs.jcim.2c01071}
}

@article{zubatyukAccurateTransferableMultitask2019,
  title = {Accurate and Transferable Multitask Prediction of Chemical Properties with an Atoms-in-Molecules Neural Network},
  author = {Zubatyuk, Roman and Smith, Justin S. and Leszczynski, Jerzy and Isayev, Olexandr},
  year = 2019,
  month = aug,
  journal = {Science Advances},
  volume = {5},
  number = {8},
  pages = {eaav6490},
  doi = {10.1126/sciadv.aav6490}
}

\appendix

\setcounter{figure}{0}
\renewcommand{\thefigure}{S\arabic{figure}}
\renewcommand{\thetable}{S\arabic{table}}
\onecolumn
\section*{Supporting Information}

\subsection*{S1: Solvation Free Energy Calculations on the FreeSolv database.}
\begin{table}[!ht]
\centering
\begin{tabular}{lrrrrrrl}
\toprule
 ID & $\Delta G_{exp}$ & $\sigma_{exp}$ & ${\Delta G_{GAFF}}$ & $\sigma_{GAFF}$ & $\Delta G_{ANI}$ & $\sigma_{ANI}$ & Selection \\
\midrule
mobley\_1328465 &  -1.99 & 0.60 &  -0.38 & 0.03 &  -1.11 & 0.04 & Random \\
mobley\_9705941 &  -1.38 & 0.60 &  -0.30 & 0.02 &  -0.26 & 0.05 & Random \\
mobley\_5494918 &   2.93 & 0.60 &   2.31 & 0.02 &   2.11 & 0.05 & Random \\
mobley\_3006808 &  -4.31 & 0.60 &  -3.81 & 0.02 &  -4.03 & 0.03 & Random \\
mobley\_8723116 &  -5.90 & 0.60 &  -5.28 & 0.03 &  -6.16 & 0.04 & Random \\
mobley\_3690931 &  -2.28 & 0.60 &  -2.75 & 0.03 &  -2.76 & 0.02 & Random \\
mobley\_1857976 &  -1.79 & 0.60 &  -0.36 & 0.02 &  -0.07 & 0.13 & Random \\
mobley\_5538249 &  -2.16 & 0.10 &  -1.15 & 0.04 &  -1.21 & 0.05 & Random \\
mobley\_9407874 &  -2.88 & 0.60 &  -2.76 & 0.03 &  -2.83 & 0.05 & Random \\
mobley\_3234716 &  -0.80 & 0.60 &  -0.66 & 0.03 &  -0.78 & 0.04 & Random \\
mobley\_4218209 &  -7.48 & 0.60 &  -7.02 & 0.03 &  -8.00 & 0.02 & Random \\
mobley\_1659169 &  -7.17 & 0.60 &  -6.12 & 0.03 &  -6.18 & 0.05 & Random \\
mobley\_9112978 &   1.31 & 0.60 &   2.37 & 0.02 &   2.40 & 0.02 & Random \\
mobley\_628951  &  -2.28 & 0.12 &  -1.23 & 0.04 &  -1.39 & 0.08 & Random \\
mobley\_2792521 &  -3.68 & 0.60 &  -3.12 & 0.03 &  -3.21 & 0.02 & Random \\
mobley\_2371092 &  -5.22 & 0.60 &  -3.20 & 0.03 &  -3.32 & 0.05 & Random \\
mobley\_967099  &  -3.13 & 0.60 &  -1.74 & 0.02 &  -1.84 & 0.03 & Random \\
mobley\_1873346 &  -0.90 & 0.20 &  -0.79 & 0.02 &  -0.79 & 0.04 & Random \\
mobley\_5072416 &  -6.12 & 0.60 &  -6.77 & 0.03 &  -7.80 & 0.02 & Random \\
mobley\_4780078 &  -6.01 & 0.60 &  -4.98 & 0.03 &  -5.10 & 0.04 & Random \\
mobley\_3047364 &  -7.65 & 0.45 & -10.55 & 0.04 & -11.31 & 0.05 & Large \\
mobley\_4587267 & -23.62 & 0.32 & -18.16 & 0.08 & -24.12 & 0.48 & Large \\
mobley\_1527293 &  -8.42 & 0.16 & -13.95 & 0.05 & -14.59 & 0.05 & Large \\
mobley\_9571888 &  -4.23 & 0.26 &  -9.78 & 0.04 & -10.07 & 0.08 & Large \\
mobley\_9534740 & -25.47 & 0.22 & -18.10 & 0.07 & -27.98 & 0.23 & Large \\
mobley\_6416775 &   2.13 & 0.60 &   2.10 & 0.03 &   2.04 & 0.06 & Small \\
mobley\_7869158 &  -1.82 & 0.10 &  -1.86 & 0.03 &  -2.12 & 0.02 & Small \\
mobley\_8337977 &  -3.05 & 0.60 &  -3.12 & 0.03 &  -3.25 & 0.04 & Small \\
mobley\_3040612 &  -0.85 & 0.10 &  -0.76 & 0.03 &  -0.98 & 0.06 & Small \\
mobley\_4375719 &   3.13 & 0.60 &   3.22 & 0.03 &   3.10 & 0.03 & Small \\
\bottomrule
\end{tabular}
\caption{Solvation free energy results for 30 selected molecules from the FreeSolv database. The table reports the molecule ID as given in the database, SFE values $\Delta G_X$ and uncertainty $\sigma_X$ for experimental and predicted values with GAFF (from database) and with ANI2x. The last column reports if the respective molecule was selected by random or large/small deviations between experimental and GAFF-predicted value. }
\label{tab:sfe_results}
\end{table}

\clearpage
\subsection*{S2: Effects of ML Region Size and Embedding on Protein-Ligand Binding}

\begin{figure}[ht!]
    \centering
    \includegraphics[width=0.7\linewidth]{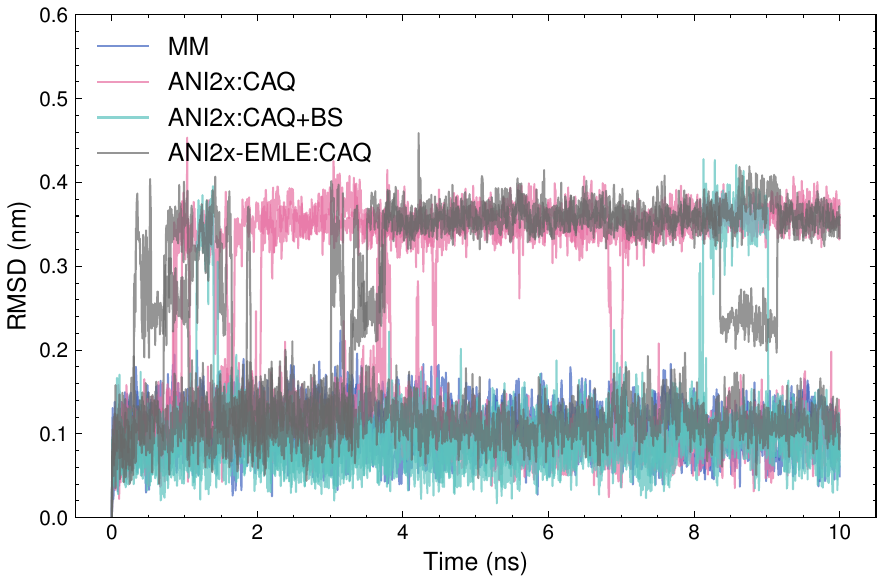}
    \caption{RMSD calculated with respect to the equilibrated starting structure for three replicates of one pure MM and three ML/MM simulations: Using ANI2x for the catechol ligand (CAQ), ligand plus binding site (CAQ+BS), and EMLE for catechol.}
    \label{fig:rmsd_all}
\end{figure}

\subsubsection*{Solvation Structure of Phenol in Water}
\begin{figure}[ht!]
    \centering
    \includegraphics[width=\linewidth]{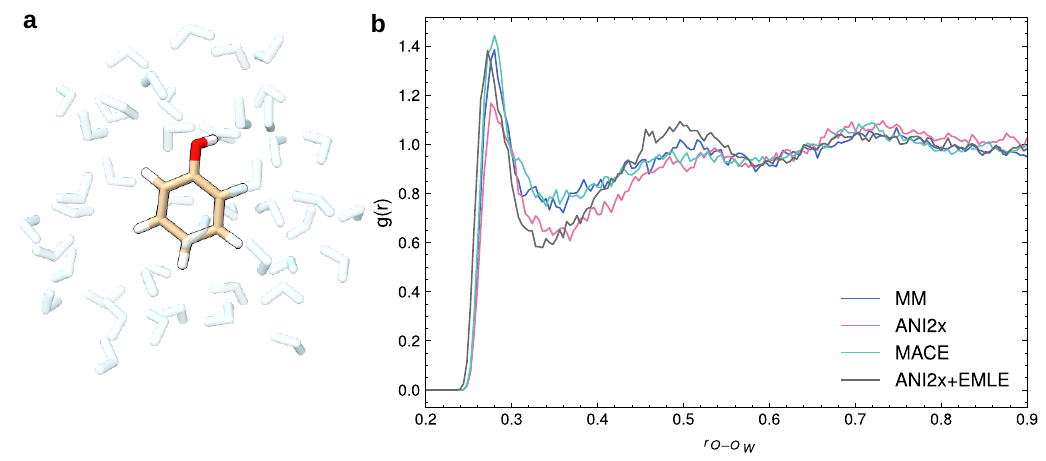}
    \caption{Solvation structure of phenol in water. (a) Structure of phenol molecule with surrounding water molecules. (b) Solvation structure of phenol in water, showing RDF between the phenol oxygen and water oxygens, calculated from 100 ps MD simulations for a pure MM description and ML/MM simulations with ANI2x, MACE and ANI2x-EMLE models, respectively.}
    \label{fig:phenol}
\end{figure}

Here, we briefly discuss the solvation structure of phenol in water. With it's relatively simple structure and hydrogen bonding OH-site, it is a good candidate to investigate also the effects of an electrostatic embedding model like EMLE \cite{zinovjevElectrostaticEmbeddingMachine2023}, and is readily comparable with previous results \cite{semelakAdvancingMultiscaleMolecular2025}. The corresponding RDF is shown in Fig. \ref{fig:phenol}. We here expand on the results in Ref. \cite{semelakAdvancingMultiscaleMolecular2025} by extending the simulation time to 200 ps, including a different base NNP model (MACE) as well as an alternative electrostatic embedding (EMLE). 

To perform the simulations, we first created a simulation box with a phenol molecule parametrized with the General Amber Force Field (GAFF) \cite{wangDevelopmentTestingGeneral2004} and solvated with 3991 TIP3P \cite{jorgensenComparisonSimplePotential1983} water molecules. Energy minimization was performed for 10000 steps, after which the system was allowed to equilibrate for 100 ps in the NVT and 100 ps in the NPT ensemble using stochastic velocity and cell rescaling \cite{bussiCanonicalSamplingVelocity2007, bernettiPressureControlUsing2020} at 300 K and 1 bar, respectively, as well as a time step of 2 fs.
Production simulations in the NVT ensemble were carried out from the final conformation of the NPT equilibration for 200 ps at a time step of 0.5 fs for pure MM, ANI-2x/MM, MACE/MM and EMLE/MM descriptions. For the calculation of the RDFs, which was performed with the \texttt{gmx rdf} command line tool, the first 5 ps were discarded from further analysis. 

It can be seen that treating the phenol molecule with the ANI-2x or MACE potential, as opposed to pure MM, does not seem to have a significant effect on the RDF. This is to be expected, as the interaction with the water molecules surrounding the OH-group is modeled with classical electrostatic and Lennard-Jones potentials in all three of these cases. \\
In the case of the EMLE model, we do see a significant departure from this behavior, as the first peak is shifted slightly to the left and sharper, a result in line with previous results \cite{semelakAdvancingMultiscaleMolecular2025}. Furthermore, we see stronger coordination in the second hydration shell, which is accessible to us because of longer simulation times. In terms of computational performance, the MM production simulation ran at 200 ns/day, ANI-2x optimized with NNPOps at 29 ns/day, MACE at 2.6 ns/day and EMLE at around 3.6 ns/day, all at a timestep of 0.5 fs. We note however, that a QM/MM simulation of the same system on comparable hardware would exhibit a performance on the order of \textit{ps/day}. In contrast, even with the slowest modules, our results were obtained within a few hours of simulation wall-time.

\clearpage
\subsection*{S3: Performance Scaling of Water Boxes}
\begin{table}[!htbp]
    \centering
    \caption{Performance benchmarks for water boxes of increasing size in terms of walltime per step (ms). Results were obtained on a NVIDIA RTX 3070 GPU at 32-bit floating point precision.}
    \label{tab:perf_ms_step}
    \begin{tabular}{
        S[table-format=5.0] 
        S[table-format=3.3] 
        S[table-format=3.3] 
        S[table-format=2.3] 
        S[table-format=2.3] 
        S[table-format=3.3] 
    }
        \toprule
        {Atoms} & {ANI-2x} & {MACE-OFF-s} & {AIMNet2} & {NNPOps} & {Nutmeg} \\
        \midrule
        3 & 8.755 & 11.253 & 7.130 & 0.973 & 9.339 \\
        9 & 8.110 & 10.925 & 7.191 & 1.018 & 9.112 \\
        21 & 8.625 & 10.883 & 7.320 & 1.611 & 10.411 \\
        30 & 8.715 & 10.871 & 7.495 & 2.094 & 10.445 \\
        90 & 9.446 & 12.034 & 7.495 & 5.226 & 13.867 \\
        210 & 10.281 & 13.850 & 7.626 & 11.621 & 24.898 \\
        300 & 9.839 & 16.768 & 9.635 & 16.377 & 33.461 \\
        900 & 10.687 & 59.991 & 70.967 & {--} & 89.567 \\
        2100 & 12.746 & 153.515 & {--} & {--} & 198.964 \\
        3000 & 16.019 & 221.271 & {--} & {--} & 297.982 \\
        9000 & 38.068 & {--} & {--} & {--} & 840.676 \\
        21000 & 90.338 & {--} & {--} & {--} & {--} \\
        30000 & 133.865 & {--} & {--} & {--} & {--} \\
        \bottomrule
    \end{tabular}
\end{table}

\begin{table}[!htbp]
    \centering
    \caption{Performance benchmarks for water boxes of increasing size in terms of throughput (ns/day) at a 1 fs time step. Results were obtained on a NVIDIA RTX 3070 GPU at 32-bit floating point precision.}
    \label{tab:perf_ns_day}
    \begin{tabular}{
        S[table-format=5.0] 
        S[table-format=2.3] 
        S[table-format=2.3] 
        S[table-format=2.3] 
        S[table-format=2.3] 
        S[table-format=2.3] 
    }
        \toprule
        {Atoms} & {ANI-2x} & {MACE-OFF-s} & {AIMNet2} & {NNPOps} & {Nutmeg} \\
        \midrule
        3 & 9.869 & 7.678 & 12.117 & 88.840 & 9.252 \\
        9 & 10.654 & 7.908 & 12.014 & 84.882 & 9.482 \\
        21 & 10.017 & 7.939 & 11.804 & 53.632 & 8.299 \\
        30 & 9.914 & 7.948 & 11.528 & 41.254 & 8.272 \\
        90 & 9.147 & 7.180 & 11.528 & 16.534 & 6.231 \\
        210 & 8.404 & 6.238 & 11.329 & 7.435 & 3.470 \\
        300 & 8.781 & 5.153 & 8.968 & 5.276 & 2.582 \\
        900 & 8.085 & 1.440 & 1.217 & {--} & 0.965 \\
        2100 & 6.779 & 0.563 & {--} & {--} & 0.434 \\
        3000 & 5.394 & 0.390 & {--} & {--} & 0.290 \\
        9000 & 2.270 & {--} & {--} & {--} & 0.103 \\
        21000 & 0.956 & {--} & {--} & {--} & {--} \\
        30000 & 0.645 & {--} & {--} & {--} & {--} \\
        \bottomrule
    \end{tabular}
\end{table}

\end{document}